\documentclass[12pt]{article}
\usepackage[top=1in, bottom =1in, left =1in, right=1in]{geometry}
\usepackage{amssymb,amsmath,cite,bm,color}
\usepackage[normalem]{ulem}
\usepackage{epsfig}
\usepackage{graphics}

\newcommand{\be}{\begin{equation}}
\newcommand{\ee}{\end{equation}}
\newcommand{\bea}{\begin{eqnarray}}
\newcommand{\eea}{\end{eqnarray}}

\newcommand{\nd}{\noindent}

\begin{document}

\title{The duality of spatial death-birth and birth-death processes and limitations of the isothermal theorem}
\author{Kamran Kaveh $^{1}$, Natalia L. Komarova $^{2}$, Mohammad Kohandel $^{1}$\\
$^{1}$Department  of Applied Mathematics, University of Waterloo, \\
Waterloo, Ontario, Canada N2L 3G1 \\
$^{2}$Department of Mathematics and Department of Ecology \& Evolutionary \\ Biology,
University of California Irvine, \\
Irvine, CA 92697, United States}

\maketitle
\begin{abstract}
Evolutionary models on graphs, as an extension of the Moran process, have two major implementations: birth-death (BD) models (or the invasion process) and death-birth (DB) models (or voter models). The isothermal theorem states that the fixation probability of mutants in a large group of graph structures (known as isothermal graphs, which include regular graphs) coincides with that for the mixed population. This result has been proven by Lieberman et al (Nature 433: 312-316, 2005) in the case of BD processes, where mutants differ from the wild types by their birth rate (and not by their death rate).  In this paper we discuss to what extent the isothermal theorem can be formulated for DB processes, proving that it only holds for mutants that differ from the wild type by their death rate (and not by their birth rate). For more general BD and DB processes with arbitrary birth and death rates of mutants, we show that the fixation probabilities of mutants are different from those obtained in the mass-action populations. We focus on spatial lattices and show that the difference between BD and DB processes on 1D and 2D lattices are non-small even for large population sizes. We support these results with a generating function approach that can be generalized to arbitrary graph structures. Finally, we discuss several biological applications of the results.
\end{abstract}

\section{Introduction}

Exploring the effect of spatial structure of an evolutionary system and its importance on the dynamics of selection has long been of interest in population genetics \cite{kimurabook}. The original stepping stone model developed by Weiss and Kimura is the backbone of the subsequent implementations of spatial structure into evolutionary modeling \cite{kimuraweiss}. Maruyama analyzed the fixation behavior of a Moran process in a geographically structured population, and was able to show that regular spatial structures do not enhance or suppress selection \cite{maruyama1,maruyama2}. More recently, Liberman et al. revisited the problem  and extended previous models to an arbitrary graph, where at each node of the graph a single species can reside \cite{lieberman}. They were able to show that some categories of graphs (such as a star graph, where a central vertex is connected to all leaf vertices) enhance the selection, while other graphs suppress the selection (see also \cite{broom,hooch}). Other notable recent works have focused on the effect of spatial structure on evolution of cooperation \cite{ohtsuki1} and on the replicator dynamics on a graph structure \cite{ohtsuki2}. Selection dynamics on various types  of random graphs has been the subject of a lot of interest recently in the context of applications to social networks \cite{redner1}, infectious disease and epidemiology \cite{keeling,britton}, and cancer modeling \cite{komarova,manem} among others. Multi-hit processes and tumour suppressor gene inactivation \cite{durret-foo}, spatial pattern formation in evolutionary models \cite{nelson,hallatschek}, and the effect of spatial distribution of fitness on selection \cite{ranfit} are some other directions of current research. For a general review of evolutionary dynamics on graphs and some recent trends see \cite{westpoint}.

In the Moran process, we assume the existence of $N$ individuals. To keep the population constant it is assumed that each elementary update consists of a birth and a death event, where individuals are chosen for birth and death from the pool consisting of the entire population. It is of little consequence whether the first event in each elementary update is a birth or a death event. However, the order of these events can become important if we consider a spatial generalization of the Moran process.

To this end, we consider a regular spatial lattice or an unstructured mesh, and again impose the condition of a constant population. For each update, an independent birth (or death) event can occur at a randomly chosen site of the lattice, but the subsequent death (or birth) event should now happen at a neighboring site of the lattice to represent the spatial structure and the fact that only daughters of individuals that are connected to a particular site can be placed there. In a birth-death (BD) process, a birth event is followed by a death event, and in a death-birth (DB) process, a death event happens first. These are the two main implementations of the generalized Moran process that have been considered in the literature.

Aside from death-birth or birth-death models there have been other implementations of evolutionary processes on a graph structure. Sood et al. \cite{redner1, redner2} discuss a model of link dynamics, where instead of random sampling of nodes, one considers random samples edges (links) of a graph as a first elementary event and then updates the values of the two nodes on the chosen link based on the fitness of individuals there (similar models are discussed in \cite{nowakbook, ohtsuki2}). Evolutionary games in the context of linking dynamics are also studied in \cite{pacheco, pacheco2, wu}.  Another variation of birth-death models on a graph is considered by \cite{vanbaalen} where each of the elementary events can be either a birth or a death or a migration (see also \cite{komarovamigration, manem}).

It has been observed by \cite{maruyama1,maruyama2} and later generalized by \cite{lieberman} that in the case of the BD process, selection dynamics are not affected by regular and symmetric structures (or more generally isothermal graphs). This elegant result is referred to as the Isothermal theorem. In this paper we explore the applicability of this theorem to other types of processes. The motivation for this question is the result of \cite{komarova} where the Moran process was studied on a 1D spatial lattice, and the DB formulation was used. It was shown that the probability of mutant fixation in this case was different from the one obtained in the space-free Moran process. Therefore, in this paper we study the connection between DB and BD processes on spatial lattices and explore the extent to which the order of birth and death events influences the probability of mutant fixation.

In the literature, differences between DB and BD updates have been described in somewhat different contexts. In papers by \cite{fu,ohtsuki3}, evolutionary games were studied on graphs, and different behavior under DB and BD models has been emphasized. In \cite{ohtsuki2}, the replicator equation on graphs is studied in the context of DB, BD, and ``imitation" dynamics; applications include evolutionary games such as  Prisoner's Dilemma, the Snow-Drift game, a coordination game, and the Rock-Scissors-Paper game. Paper \cite{zukewich} studied evolution of cooperation, and formulates the corresponding Moran process under the DB and BD implementations.  It is shown that for DB updates, cooperation may be favored in structured populations, while with BD updates this never is. The authors propose a mixed rule where in each time step DB or BD updates are used with fixed probabilities, and they further derive the conditions for selection favoring cooperation under the mixed rule for various social dilemmas. Our work adds to these investigations by studying the difference between DB and BD updates in the Moran process in the context of mutant fixation.

Another natural generalization of the Moran process that we focus on here, concerns the definition of fitness. In the conventional Moran model, all individuals are characterized by their ``fitness" parameters. Usually, an individual is chosen for division with a probability weighted by its fitness, and an individual is chosen for death with the probability $1/N$ (such that all individuals are equally likely to be chosen for death, see e.g. \cite{lieberman,zukewich}). In a more general setting, however, fitness could be influenced by the death rate as much as it is by the division rate. For example, a cancer cell in tissue could differ from the surrounding  cells by its division rate and/or by its death rate. Therefore, in this paper we aim to explore how the division and the death rates both influence the fixation dynamics of mutants in the DB and BD processes.

The rest of this paper is organized as follow. In section \ref{sect:def} we formulate  DB and BD models on a graph, with general death and birth rates. We show that  DB and BD transition probabilities can be transformed into each other by using a duality property of the two models. To study the differences between the DB and BD processes, we start with the case of a complete graph (where all vertices are connected, section \ref{sect:compl}) and derive closed algebraic forms for fixation probability in either of cases. In particular, we show that  the difference between DB and BD implementation on a complete graph vanishes as $1/N$, as the system size goes to infinity. In section \ref{sect:1D} we discuss exact solutions of DB and BD models on a 1D circle and show that the differences are no longer negligible even for large system sizes. In section \ref{sect:2D} we find an approximation for the fixation probability in higher dimensions and the higher connectivities. Based on the results obtained, we conjecture that the difference between the two models has the order of the inverse of the degree of connectivity of the graph. This is supported by the exact solutions in the complete graph and 1D cases, and numerical simulations in the 2D case. We show that our approximation in 2D for a regular lattice matches the exact stochastic simulations within a very good error margin. Finally, a discussion is presented in section \ref{sect:disc}.

\section{Death-Birth (DB) and Birth-Death (BD) processes on a graph}
\label{sect:def}

Assume two species, A (normal, or wild type) and B (mutant), with corresponding proliferation rates $r_{\rm B}$ and $r_{\rm A}$ and death rates $d_{\rm B}$ and $d_{\rm A}$. A general death-birth process (DB) is defined as follows: at each time step, one cell at site $i$ (either A or B) is randomly chosen to die with weight $d_{\rm A}$ or $d_{\rm B}$. One of the neighbouring sites to $i$ (denoted by $j$) is chosen at random, with weight $r_{\rm A}$ or $r_{\rm B}$ to proliferate, and probability $w_{ij}$ for the offspring to be placed in site $i$ (to replace the dead cell). For each site $i$, we set $n_i=1$ if the resident cell is mutant, and $n_i=0$ if the resident cell is wild type. Thus the distribution of mutant cells at sites $i$ is $n_{i}$ and the distribution of normal cells is $1 - n_{i}$. Given the vector $\vec{n} = (n_{1},..,n_{N})$, the transition probabilities can be written as:

\bea
W^{+}_{\rm DB}(\{n_{i}\}) 
&=& \frac{d_{\rm A}(1- n_{i})}{d_{\rm B}\sum_{k}n_{k} + d_{\rm A}\sum_{k}(1- n_{k})}\frac{r_{B}\sum_{j}w_{ij} n_{j}}{(r_{\rm B}-r_{\rm A})\sum_{l}w_{il}n_{l} + r_{\rm A}} \nonumber\\
W^{-}_{\rm DB}(\{n_{i}\})
&=& \frac{d_{\rm B}n_{i}}{d_{\rm B}\sum_{k}n_{k} + d_{\rm A}\sum_{k}(1- n_{k})}\frac{r_{\rm A}\sum_{j}w_{ij} (1-n_{j})}{(r_{\rm B}-r_{\rm A})\sum_{l}w_{il}n_{l} + r_{\rm A}}.
\label{eq1}
\eea
Here, $W_{\rm DB}^+$ ($W_{\rm DB}^-$) stands for the probability that in the death-birth process, a new mutant (a new wild-type cell) appears at site $i$ after one elementary update. Similarly, in a birth-death process (BD) the sequence of the two death and birth events is switched and can formally be obtained by switching $r$'s and $d$'s in equation~(\ref{eq1}), and also $w_{ij} \rightarrow w_{ji}$.

\bea
W^{+}_{\rm BD}(\{n_{i}\}) = \frac{r_{\rm B}n_{i}}{r_{\rm B}\sum_{k}n_{k} + r_{\rm A}\sum_{k}(1- n_{k})}\frac{d_{\rm A}\sum_{j}w_{ji} (1-n_{j})}{(d_{\rm B}-d_{\rm A})\sum_{l}w_{li}n_{l} + d_{\rm A}}\nonumber\\
W^{-}_{\rm BD}(\{n_{i}\}) = \frac{r_{\rm A}(1- n_{i})}{r_{\rm B}\sum_{k}n_{k} + r_{\rm A}\sum_{k}(1- n_{k})}\frac{d_{B}\sum_{j}w_{ji} n_{j}}{(d_{\rm B}-d_{\rm A})\sum_{l}w_{li}n_{l} + d_{\rm A}}.
\label{eq2}
\eea

In the case of BD, $W^{+}_{\rm BD}(\{n_{i}\})$ indicates that a new mutant appears at any site $j$ neighbouring site $i$. Notice that in the case of a DB process, the death event is a {\it global} event and the individual is chosen for death among all the individuals
on the graph with weight of $d_{\rm A}$ or $d_{\rm B}$. The following birth event, however, takes places among only the individuals
that are connected to the previously chosen individual,  and thus constitutes a {\it local} event. Even though both of the events occur through random sampling with weighted biases $d_i$ or $r_i$, with $i={\rm A,B}$, the subset of random sampling  is smaller for the second event unless the graph is a complete graph. Similarly, in a BD process, the global event is a birth event, while the death event takes place for the neighbours of the chosen individual on the graph and thus is a local event.

There are specific cases of the above processes that have been often discussed in the literature. The case of $d_{\rm A} = d_{\rm B} = 1$ has been typically implemented as either a DB or BD process on a graph. As the reader can easily see, the opposite limit where $r_{\rm A} = r_{\rm B} = 1$ while $d_{\rm A,B}$ are arbitrary, also leads to equivalent models while the dynamics of the two models are reversed. In other words a DB model with $r_{\rm A}=r_{\rm B} = 1$ is equivalent to a BD model with $d_{\rm A} = d_{\rm B} = 1$ and vice versa.

In the case of BD models with $d_{\rm A} = d_{\rm B}$ it has been argued that for a wide range of graph structures, known as isothermal graphs, where $\sum_{j} w_{ij} = \sum_{i} w_{ij} = 1$, the fixation probability of the model is the same as the mixed population one \cite{lieberman} (see also \cite{maruyama1}, \cite{maruyama2} and \cite{hooch}).
The main examples of isothermal graphs are regular lattices with periodic boundary conditions, and the introduction of boundaries or
randomness in connectivities removes the isothermal property (for effect of boundaries see \cite{komarova},\cite{manem} and \cite{broom}).

\section{Fixation probability for DB and BD processes on a complete graph}
\label{sect:compl}

Let us consider the simplest process which is on a complete graph, where every two nodes $i$ and $j$ are connected, i.e. $w_{ij} = 1/N$. Let us use the following convenient notation:
$$r_B/r_A=r,\quad d_B/d_A=d.$$

The Kolmogorov backward equation for the absorption probability $\pi_{m}$, where $m$ is the mutant population, is written as
\bea
\pi_{m} &=& W^{+}_{m+1}\pi_{m+1} + W^{-}_{m-1}\pi_{m-1} + (1 - W^{+}_{m} - W^{-}_{m})\pi_{m},\nonumber\\
\pi_{0} &=& 0, ~~~~~~ \pi_{N} = 1,
\label{eq3}
\eea

\nd which is valid for both BD and DB processes. Transition probabilities $W^{\pm}_{m}$ are defined as the probabilities to gain or lose one mutant in a system with $m$  mutants. The general solution for the fixation probability starting from only one mutant can be found in closed form. Denoting
$$\gamma_m=\frac{W_m^-}{W_m^+},$$
we obtain for the probability of fixation,
\be
\pi_{1} = \displaystyle \frac{1}{\displaystyle 1 + \sum^{N-1}_{j=1}\gamma_1\ldots\gamma_j}.
\label{eq4}
\ee
Notice that the above result is true when the transition probabilities in such a one step process only depend on the number of mutants in the system at every time step, and not on other degrees of freedom of the system. This condition holds for the complete graph, and, as we will see later, for 1D rings.

Next, we will consider some implementations of DB and BD processes on a complete graph. First, we will assume the  update rules where the second event (death in the BD process and division in the DB process) occur in a neighborhood of a given cell, which includes the cell itself. That is, for example, if a cell is chosen for reproduction first, during the death event this (mother) cell will be in a pool of cells that have a probability to die. This is equivalent to having a graph on which every node is connected to itself by a loop. (Ohtsuki and Nowak \cite{ohtsuki1} call this an "imitation" updating in the context of game theory on graphs.)

Later in this section, we will  include a process where the neighborhood does not includes ``self'' as a candidate for the second event to take place. The preference for either of the two updatings is somewhat arbitrary and depends on the particular nature of the modeling problem at hand.

\subsection{The complete network including ``self"}

First, we will consider the process where the network of neighbors of a given cell includes the cell itself. To explain the update procedure in more detail, note that at every time step, we randomly {\it label} a cell for death (birth) and then label a cell for birth (death). We let the death and birth events happen after both labels have been assigned. This way, for the complete graph (mass-action) scenario, the number of ``neighbors" is always $N$, and we have the following transition probabilities:
\bea
W^{+}_{m} = \frac{rm}{rm + N-m}\cdot\frac{N-m}{dm +N-m}\nonumber\\
W^{-}_{m} = \frac{N-m}{rm + N-m}\cdot\frac{dm}{dm +N-m}.
\eea
Interestingly, these probabilities are the same for the DB and BD processes, the two are completely equivalent in this case. We therefore have
$$\gamma_m=\frac{d}{r},$$
and the following well-known result for the fixation probability holds:
\be
\label{mor}
\pi_1=\frac{1-d/r}{1-(d/r)^N}.
\ee

Next, we will study networks that do not include ``self''. For example, in the BD process described below, after the initial birth event, the cell that has just divided is excluded from the death event. In the following section, for the DB process, the cell that has been chosen for death will not be participating in the subsequent reproduction event.

\subsection{The BD process}

For a BD process on a complete graph excluding ``self'', the transition probabilities $W^{\pm}_{m}$ are written as,

\bea
W^{+}_{{\rm BD},m} = \frac{rm}{rm + N-m}\cdot\frac{N-m}{dm +N-m -d}\nonumber\\
W^{-}_{{\rm BD},m} = \frac{N-m}{rm + N-m}\cdot\frac{dm}{dm +N-m-1}.
\label{WBD}
\eea

For $d=1$ the ratio of $W^{-}_{{\rm BD},m}/W^{+}_{{\rm BD},m} = 1/r$ for any $m$, however, this is not in general true as upon choosing a mutant/normal cell to divide (birth event) the next death event among the rest of the $N-1$ cells will normalize differently in $W^{-}$ and in $W^{+}$. In other words, when a mutant cell is chosen to divide ($W^{+}$ contribution) the successive death event happens among the rest of the $m-1$ mutants (excluding the one already chosen to divide) and $N-m$ existing normal cells. When a normal cell is chosen to divide ($W^{-}$ contribution) the successive death event occurs among $m$ mutant and $N-m-1$ normal cells. In general the ratio of $W^{-}_{{\rm BD},m}/W^{+}_{{\rm BD},m}$ is,

\be
\gamma^{\rm BD}_{m}\equiv \frac{W^{-}_{{\rm BD},m}}{W^{+}_{{\rm BD},m}} = \frac{d}{r}\times \left( 1 - \frac{d-1}{N-m+dm-1}\right).
\label{gammaBD}
\ee

For $d=1$ the ratio will be $1/r$ and substituting into equation~(\ref{eq4}) give rise to well-known Moran result, equation~(\ref{mor}),

\be
\pi_{1} = \frac{\displaystyle 1 - \frac{1}{r}}{\displaystyle 1 - \left(\frac{1}{r}\right)^{N}}.
\label{piA}
\ee

For general values of $d$, however, a closed form for $\pi_{1}$ can be obtained,

$$\pi_1^{\rm BD}=\frac{(d-1)r}{d(N-1)\Phi\left(\frac{d}{r},1,\frac{N+d-2}{d-1}\right)+r\left(d-1-(N-1)\left(\frac{d}{r}\right)^N\Phi\left(\frac{d}{r},1,\frac{N+d-2}{d-1}\right)\right)} ,$$
where the Lerch transcendent is defined as
$$\Phi(z,s,a)=\sum_{k=0}^\infty\frac{z^k}{(a+k)^s}.$$
To evaluate the large-$N$ behavior of the fixation probability for advantageous mutants, we note that the function $\Phi\left(\frac{d}{r},1,\frac{N+d-2}{d-1}\right)$ decays as a power law as $N\to\infty$, and $(d/r)^N$ decays exponentially for $d<r$. Therefore, we can ignore the term multiplying $(d/r)^N$. Further, we notice the following asymptotic behavior of the Lerch transcendent:
$$\Phi(z,1,a)=\sum_{k=0}^\infty \left(\frac{z^k}{a}+\frac{kz^k}{a^2}\right)+O(1/a^3)=\frac{1}{a(1-z)}+\frac{z}{a^2(1-z)^2}+O(1/a^3).$$
Substituting this expression with $z=d/r$ and $a=\frac{N+d-2}{d-1}$ into the equation for $\pi_1$, and expanding further in Taylor series in $1/N$, we obtain
\be
\label{piBDapp}
\pi_1^{\rm BD}=1-\frac{d}{r}+\frac{(d-1)d}{rN}+O(1/N^2).
\ee
The same expansion can also be obtained by a different method. If we expand the expression for $\gamma_m^{BD}$, equation~(\ref{gammaBD}), in terms of small $1/N$, we obtain
\be
\label{BDappr}
\gamma_m^{\rm BD}=\frac{d}{r}-\frac{(d-1)d}{rN}+O(1/N^2),
\ee
an expression that does not depend on $m$. This means that up to the order $O(1/N)$ we can simply use the geometric progression formula and obtain the result presented in equation~(\ref{piBDapp}).

\subsection{The DB process}

Similarly, for a DB process on a complete graph, the transition probabilities $W^{\pm}_{m}$ are given by,

\bea
W^{+}_{{\rm BD},m} = \frac{N-m}{dm + N-m}\cdot\frac{rm}{rm+N-m-1}\nonumber\\
W^{-}_{{\rm DB},m} = \frac{dm}{dm + N-m}\cdot\frac{N-m}{rm +N-m-r},
\label{WDB}
\eea
and thus,

\be
\gamma^{\rm DB}_{m} = \frac{d}{r}\times\left(1 + \frac{r-1}{rm + N-m-r}\right).
\label{gammaDB}
\ee
Comparing equations~(\ref{gammaBD}) and (\ref{gammaDB}) one can see a duality between the two models which comes naturally from the definiton of the models,

\be
\gamma^{\rm DB}_{m}(r, d) = \frac{1}{\gamma^{\rm BD}_{m}(d,r)}.
\ee
This results in an interesting connection between the DB and BD
results for fixation probability.

Also, the fixation probability is obtained by

\be
\pi_{1}^{\rm DB} = \frac{(N-1)(d-r)^{2}}{\displaystyle\left(\frac{d}{r}\right)^{N}\left(Ndr^{2}-Nr^{3}-dr^{2}+r^{2}\right) + r(r(N-1)+d(r-N))},
\label{fixDB}
\ee
For $r=1$ this reduces to equation~(\ref{mor}),
\be
\frac{d-1}{d^{N}-1}.
\ee

The expansion of equation~(\ref{fixDB}) is,

\bea
\label{DBappr}
\pi_{1}^{\rm DB} &=& \displaystyle 1-\frac{d}{r}\nonumber\\
&+& \left(\frac{1}{r}  - 1\right)\cdot \frac{d}{N}+ \mathcal{O}\left(\frac{1}{N^{2}}\right).
\eea
Again, the same result can be obtained by first expanding the expression for $\gamma_m^{\rm DB}$, equation~(\ref{gammaDB}), in terms of $1/N$, and then performing the summation of a geometric series.

\subsection{Conclusions for the complete graph scenario}
\begin{itemize}
\item In the case where ``self" is a part of the neighborhood graph, the BD and DB processes are identical, and we obtain the exact formula for the probability of fixation, equation~(\ref{mor}).
\item If ``self" is not included, then the BD and DB processes are different from each other. The difference is of the order of $1/N$.
\item If $d=1$ for the BD process then the probability of fixation is given by expression (\ref{mor}). Similarly, if $r=1$ in the DB process, equation~(\ref{mor}) holds. These results can be interpreted as the isothermal theorem for the mass action scenario.
\item For large values of $N$, the probabilities of fixation can be approximated by expressions (\ref{BDappr}) and (\ref{DBappr}) for the BD and DB processes, respectively.

\end{itemize}

\section{Exact results for fixation probability in 1D}
\label{sect:1D}

In this section we focus on the solutions of the DB and BD models on a circle with arbitrary birth and death rates $r$ and $d$ for the mutant cells. Similar but more restricted cases have been discussed in the literature. Papers \cite{redner1},\cite{redner2} and \cite{hooch} considered a representation of the model which is a weak-selection limit of the model we discussed in the previous section and showed that the result agrees with the isothermal theorem. Paper \cite{komarova} has approached the problem more directly by solving a backward Kolmogorov equation with periodic boundary conditions. We will consider a similar implementation to \cite{komarova} and suggest an alternative and more intuitive method to calculate the fixation probability that can be extended to higher dimensions and different connectivities.

\begin{figure}[!h]
\begin{center}
\epsfig{figure=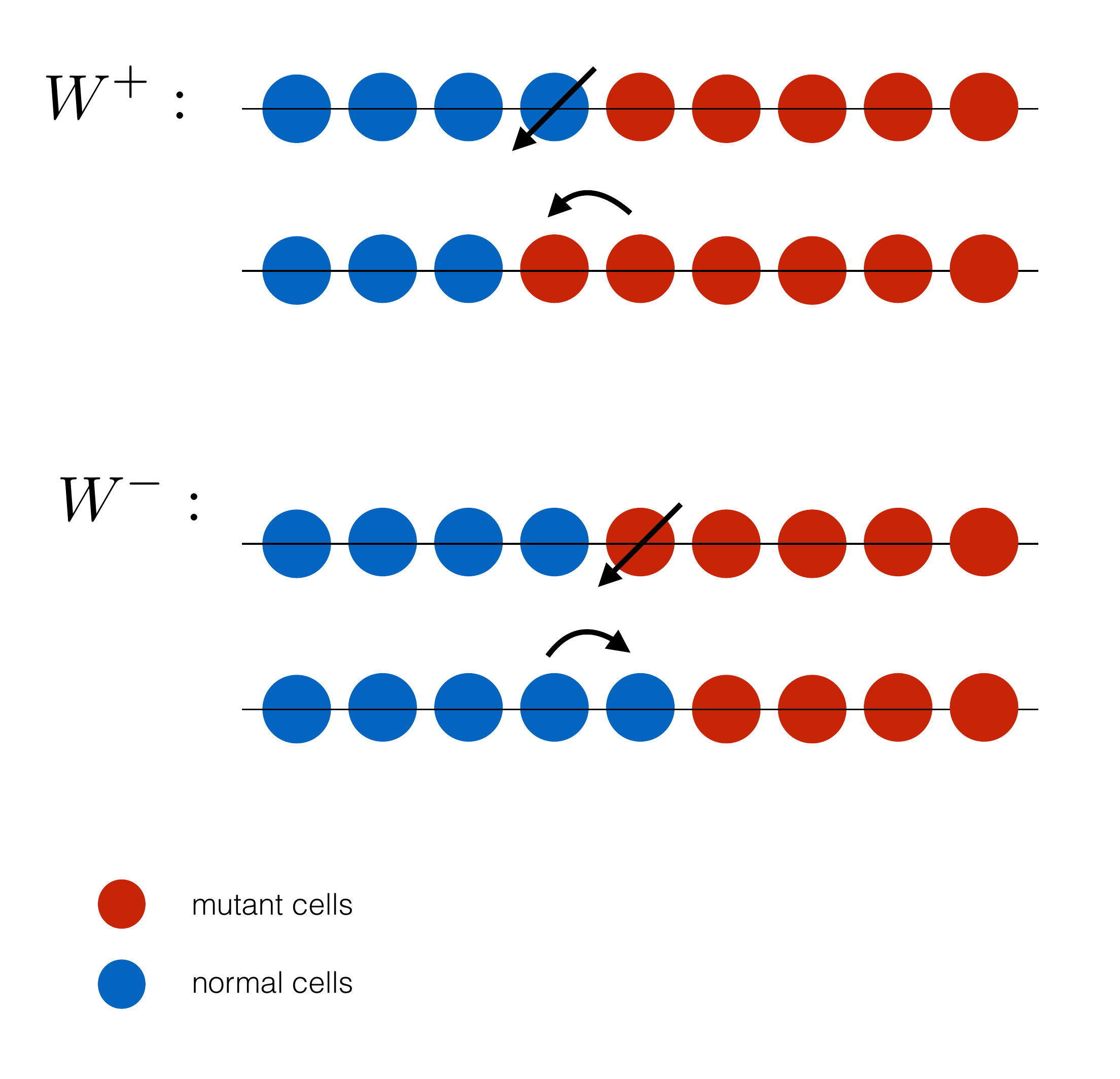, height=200pt,
width=200pt,angle=0}
\end{center}
\caption{Sequence of events in a death-birth process on a line (circle). The transition probabilities $W^{\pm}$ are indicated by the corresponding events.}
\label{voter1}
\end{figure}

\begin{figure}[!h]
\begin{center}
\epsfig{figure=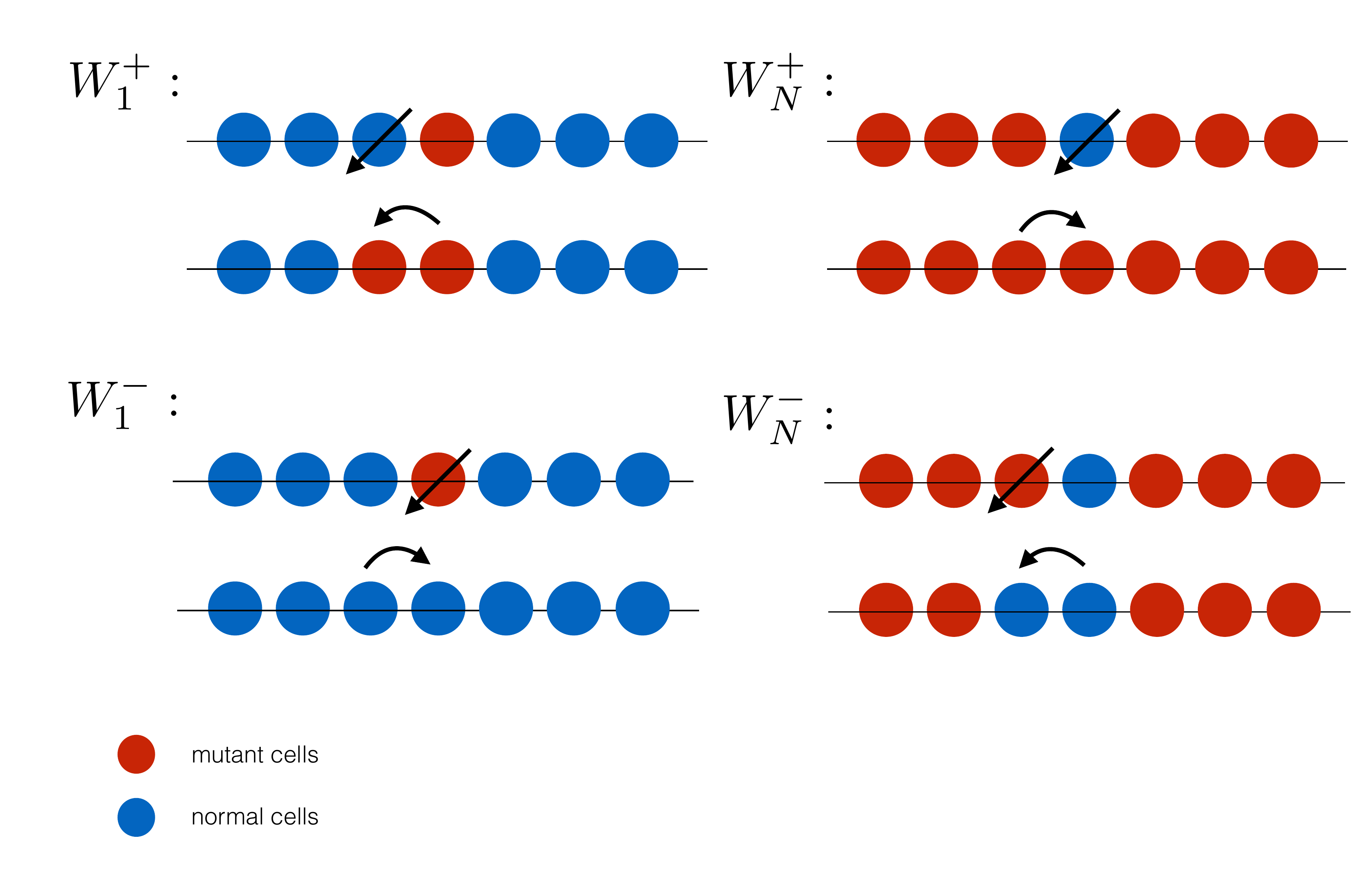, height=200pt,
width=270pt,angle=0}
\end{center}
\caption{Transition probabilities for the first and last events of fixation. In the text $\gamma_{1} = W_{1}^{-}/W_{1}^{+}$ and
$\gamma_{N-1}=W^{-}_{N-1}/W^{+}_{N-1}$. (Figure is depicted for a DB process).}
\label{voter1-1}
\end{figure}

One dimension is a particular case for which the mutant clone keeps a constant number of normal (or mutant) neighbours. Moreover, the death-birth events which lead to an increase or decrease of the mutant population only occurs at the two ends of the boundary between mutant and normal clones. On the boundary, at any time, the number of mutant neighbours and normal neighbours remain the same. The sequence of events is depicted in figure~\ref{voter1}. There are two exceptions to this condition and that is when there is one mutant and $N-1$ normal cells in the system, or $N-1$ mutant and 1 normal cell, in which the mutant (normal) cell has two normal (mutant), adjacent neighbouring cells (see figure~\ref{voter1-1}). This is not true in higher dimensions. In higher dimensions, the clonal front geometry and even the topology of a mutant clone can fluctuate.

In the following subsections we will consider the DB and BD  processes where the neighborhoods do {\it not} include ``self''. The case where ``self'' is included is considered in Appendix A.

\subsection{The DB model}

For the one-dimensional DB model, it turns out that  the condition $W^{-}_{m}/W^{+}_{m} = d/r$ holds true  for $ 1< m < N -1$, since in this case, the  transition probabilities $W^{\pm}$ can be rewritten as,

\bea
W^{+}_{\rm DB-1D,m} &=& \frac{1}{m+(N-m)d}\,\frac{2r}{r + 1}\nonumber\\
W^{-}_{\rm DB-1D,m} &=& \frac{d}{m+(N-m)d}\frac{2}{r + 1}.
\label{eq6p}
\eea

The ratio of transition probabilities is different for $m = 1$, i.e. where there is only one mutant in the population, and for $m=N-1$, where only one normal cell is left in the population. This is due to the fact that the condition of having the same number of mutant or normal cell neighbours on the boundary is not correct in the latter two cases. A single mutant cell in the beginning has no neighbouring mutant cells, and the last remaining normal cell before a full takeover does not have any normal neighbours left. Therefore, we have
\bea
\frac{W^{-}_{\rm DB-1D, m}}{W^{+}_{\rm DB-1D, m}} = \displaystyle
\left\{ \begin{array}{cc}
d/r &  1 < m < N-1\\
d(r + 1)/2r  \equiv \gamma_{1}& m = 1\\
2d/(r+1) \equiv \gamma_{N-1}& m = N - 1
\end{array}
\right.
\label{eq7}
\eea
Substituting these values into equation~(\ref{eq4}) gives,

\be
\pi^{\rm DB-1D}_{1} =\frac{2d(r-d)}{(d/r)^Nr^2(r-1-2d)+d(d(r-1)+2r)}
\label{1Dfix}
\ee
In the case where $d=1$, we obtain
$$\pi^{\rm DB-1D}_{1} =\frac{2(r-1)}{3r-1+(r-3)(1/r)^{N-2}}.$$
In the limit $N\to \infty$ and $r>1$, this reduces to
$$\pi^{\rm DB-1D}_{1} =\frac{2(r-1)}{3r-1},$$
a result obtained by \cite{komarova}. For $r=1$, equation~(\ref{1Dfix}) gives
$$\pi^{\rm DB-1D}_{1} =\frac{1-d}{1-d^N},$$
which coincides with equation~(\ref{mor}) with $r=1$. Equation~(\ref{1Dfix}) can be also obtained using a generating function method (Appendix C).

\subsection{The BD process}

Due to duality between the formulation of a DB process and its corresponding BD process (see equations~(\ref{eq1}) and (\ref{eq2})) the same calculation
can be repeated for a BD process by switching death and birth rates
($d \leftrightarrow r$ and $W^{+}\leftrightarrow W^{-}$). The transition probabilities for the corresponding BD process are,
\bea
\frac{W^{-}_{\rm BD-1D, m}}{W^{+}_{\rm BD-1D, m}} = \displaystyle
\left\{ \begin{array}{cc}
d/r &  1 < m < N-1\\
2d/r(d+1) \equiv \gamma_1 & m = 1\\
(d+1)/(2r)\equiv \gamma_{N-1} & m = N - 1
\end{array}
\right.
\label{eq11-1}
\eea
Substituting these values into equation~(\ref{eq4}) gives,
\be
\pi^{\rm BD-1D}_{1} =
\frac{d^2(1+d)(d-r)}{d^2(d(d-1)-r(d+1))+(d/r)^Nr(d(r+d+1)-r)}.
\label{eq12}
\ee
In the limit where $d=1$, the fixation probability above takes the well-known form for a Moran process,
\be
\pi^{\rm BD-1D} = \displaystyle\frac{ \displaystyle1 - \frac{1}{r}}{\displaystyle 1 - \left(\frac{1}{r}\right)^{N}},
\label{eq13-1}
\ee
see equation~(\ref{mor}) with $d=1$.

\subsection{Comparison of one-dimensional DB and BD processes}

Below we analyze the fixation probabilities for the one-dimensional DB and BD processes obtained in this section, see equations~(\ref{1Dfix}) and (\ref{eq12}).

\begin{figure}[!h]
\begin{center}
\epsfig{figure=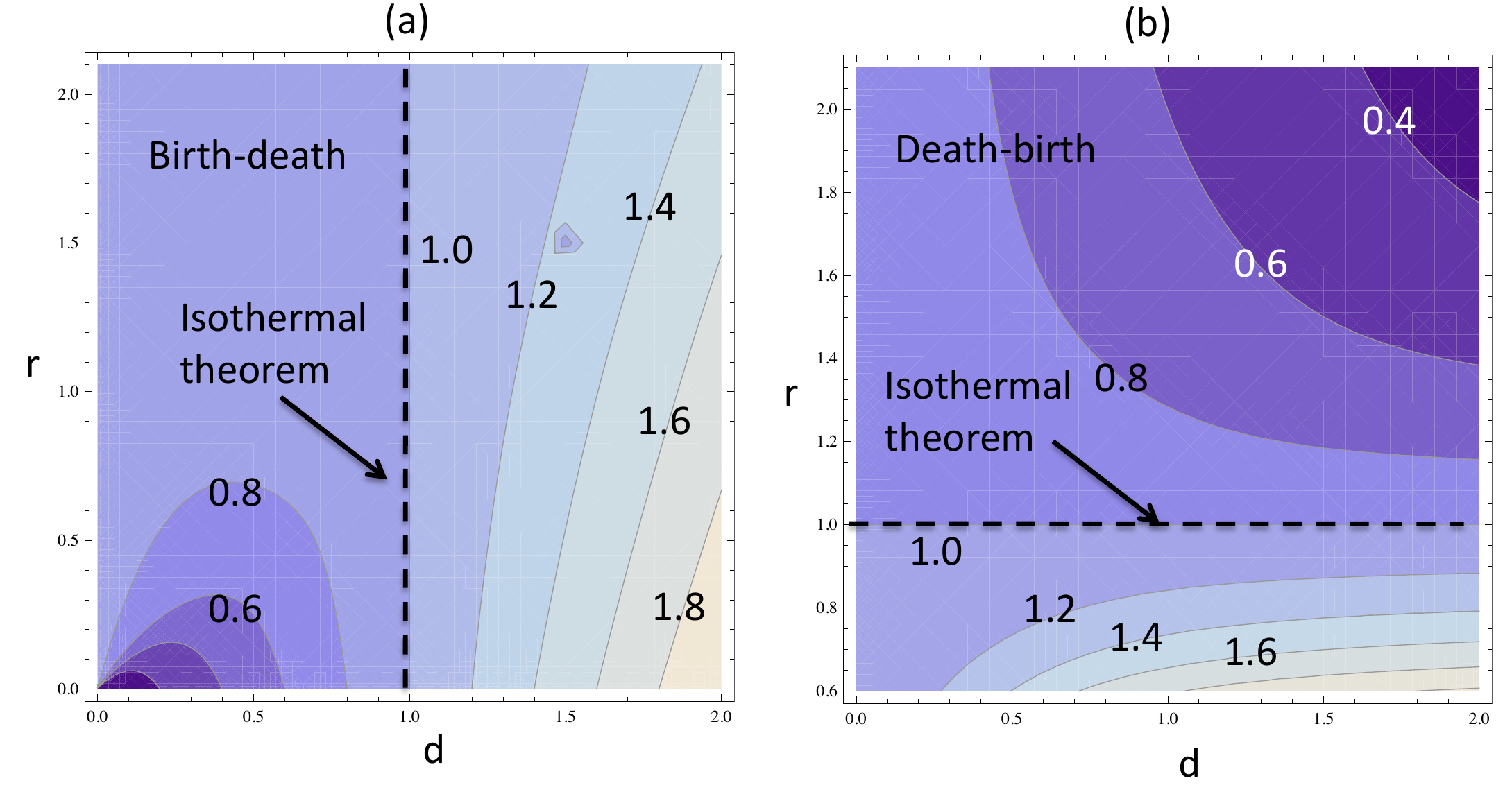, height=200pt,
width=370pt,angle=0}
\end{center}
\caption{1D process: Comparison of the BD and DB processes with the nonspatial process. (a) The ratio $\pi^{BD-1D}/\pi^{Moran}$ is plotted as a function of variables $d$ and $r$. The contours correspond to equal values of this quantity, and lighter colors mark higher values. (b) The same, for the function $\pi^{DB-1D}/\pi^{Moran}$. The parameter $N=100$ is used.}
\label{fig:BD2}
\end{figure}

\paragraph{Comparison between the 1D spatial processes and the nonspatial Moran process}
We can compare the DB and DB processes with the nonspatial Moran process for general values of $r$ and $d$. The results are as follows:
\begin{itemize}
\item If $d>1$, the BD process has a higher probability of fixation than the Moran process; if $d<1$, then it has a lower probability of fixation, see figure \ref{fig:BD2}(a).
\item If $r>1$, the DB process has a lower probability of fixation than the Moran process; if $r<1$, then it has a higher probability of fixation, see figure \ref{fig:BD2}(b).
\end{itemize}
In figure \ref{fig:BD2}, we plot the ratio of the probability of fixation in the spatial process (BD is (a) and DB in (b)) and the nonspatial Moran process. The set of values that satisfy the isothermal theorem separates the regions where the spatial process yields a higher and a lower probability of fixation, compared with the Moran process.

To prove the results above, we consider the cases $d<r$ and $r<d$ separately, and take the limit $N\to\infty$. For the case of the BD process, the expression $\pi^{BD-1D}/\pi^{Moran}$ simplifies to
$$\frac{r+dr}{d-d^2+r+dr}\quad \mbox{and} \quad \frac{d^2(1+d)}{d(1+d+r)-r}$$
in the two cases. Both expressions are greater than $1$ if $d>1$, and smaller than one otherwise.

Similarly, for the DB process, in the cases where $d<r$ and $r<d$, the ratio $\pi^{DB-1D}/\pi^{Moran}$ simplifies to
$$\frac{2r}{d-r(2+d)}\quad \mbox{and} \quad \frac{2d}{r(1+2d-r)}.$$
Both expressions are greater than one if $r<1$, and they are smaller than one otherwise.

The isothermal theorem can be considered a  ``borderline'' case, see the dashed lines in figure \ref{fig:BD2}. This can be summarized by  the following statements:
\begin{itemize}
\item The BD process satisfies the isothermal theorem in the particular case where $d=1$.
\item The DB process satisfies the isothermal theorem if $r=1$.
\end{itemize}

\paragraph{Comparing the BD and DB processes to each other}
For the same values of $d$ and $r$, the DB and BD process yield different probabilities of fixation. In the special case where $d=1$, the BD process has a higher fixation probability than the DB process iff $r>1$. That is, in the BD process, advantageous mutants have a higher probability of fixation, and disadvantageous mutants have a lower probability of fixation.

If $r>d$ and $N\to\infty$, the probabilities of fixation have a simpler form:
$$\pi^{\rm BD-1D}=\frac{(1+d)(d-r)}{d(d-1)-r(d+1)},\quad \pi^{\rm DB-1D}=\frac{2(r-d)}{d(r-1)+2r}.$$
In this limit, the two processes have the same probability of fixation if
$$r=\frac{3-d}{1+d}.$$

\begin{figure}[!h]
\begin{center}
\epsfig{figure=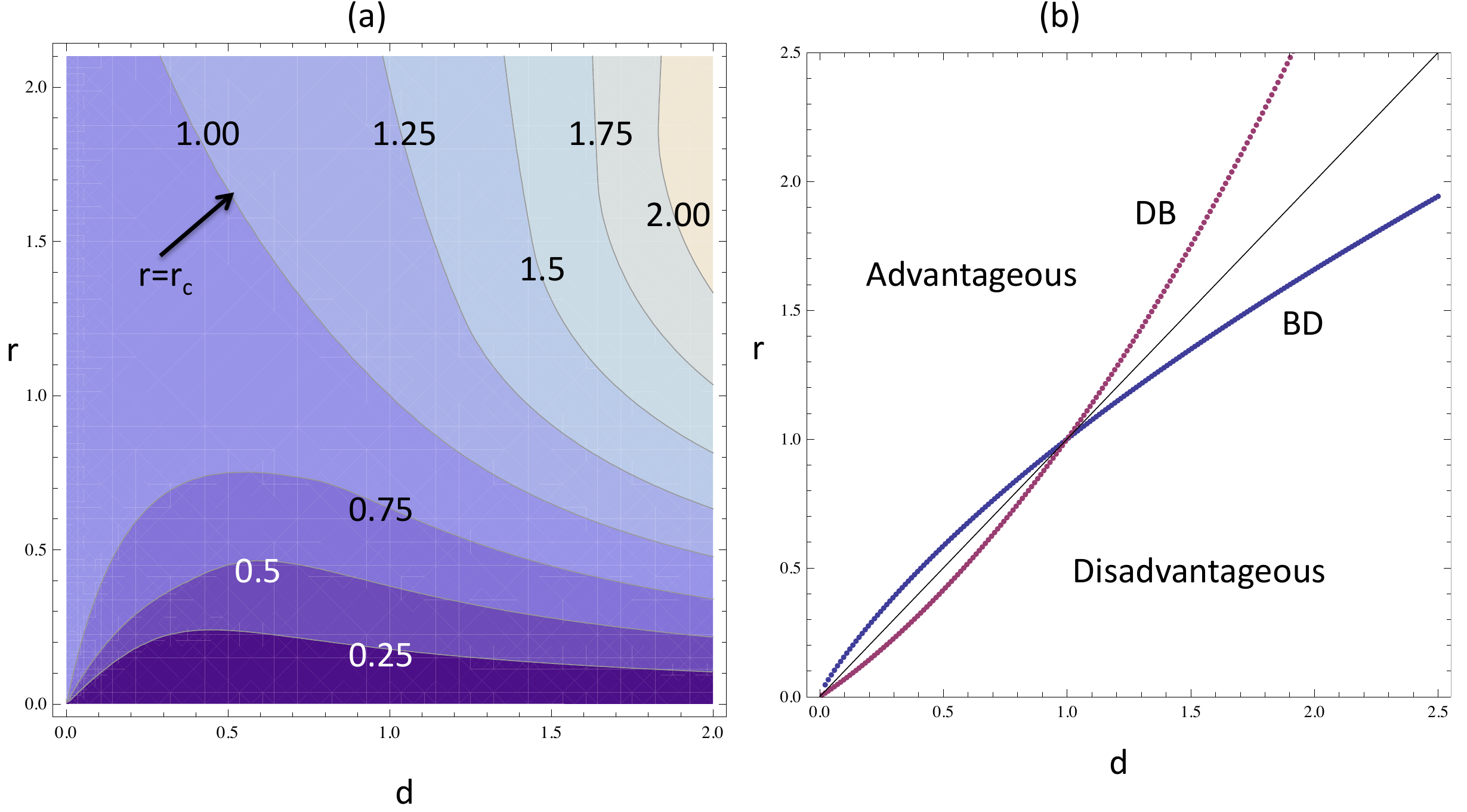, height=200pt,
 width=370pt,angle=0}
\end{center}
\caption{1D process: properties of the fixation probabilities. (a) The ratio $\pi^{BD-1D}/\pi^{DB-1D}$ is plotted as a function of variables $d$ and $r$. The contours correspond to equal values of this quantity, and lighter colors mark higher values. Above the line $r=r_c$, the probability of fixation is larger for the BD process. The parameter $N=100$ is used. (b) Neutrality of mutants. Plotted are he lines in the $(d,r)$ space along which the probability of mutant fixation is equal to $1/N$ (the line $\pi^{BD-1D}=1/N$ is blue, and the line $\pi^{DB-1D}=1/N$ is red; the line $r/d=1$ is black). Parameter $N=5$ was used. }
\label{fig:BD1}
\end{figure}

In general, there is a certain threshold value, $r_c$, such that for $r>r_c$ ($r<r_c$), the BD process (the DB process) has a higher probability of fixation. This can be interpreted as follows: for mutants that are sufficiently advantageous, the BD process is more successful, and for less advantageous mutants, the DB process is more successful. This is illustrated in figure \ref{fig:BD1}(a), where we plot the ratio $\pi^{\rm BD-1D}/\pi^{\rm DB-1D}$ as a function of the variables $d$ and $r$. The contours correspond to equal values of this quantity, and lighter colors mark higher values. Above the line $r=r_c$, the probability of fixation is larger for the BD process.

Another interesting limit is for large-$N$ and large-$r$. Even in the simple case of $d=1$, while the BD process leads to $\pi^{\rm BD}_{1} = 1 -1/r \rightarrow 1$, the DB result is $\pi^{\rm DB}_{1} = 2(r-1)/(3r-1)\rightarrow 2/3$. The fact that in a DB process, the fixation probability of an arbitrarily advantageous mutant never approaches unity might sound surprising but in fact is very intuitive. In a DB updating scheme, no matter how high the chances are for a single advantageous mutant to proliferate, the chance of extinction is always contributed by the early death events. In the case of a 1D cycle, there is a $1/N$ chance for the mutant to be picked and replaced by one of its neighbours in the first time step, while there is $2/N$ chance for the mutants neighbours to be chosen to die and be replaced by a mutant offspring. The ratio of the total number of events that the population changes to the ones that the mutant survives is $2/3$, which is exactly the fixation probability for $r \rightarrow \infty$ limit.

\paragraph{Neutrality of the mutants} The spatial processes require a different definition of neutrality for the mutant. In the nonspatial Moran process, neutral mutants satisfy $r/d=1$, and the probability of fixation of such mutants is $1/N$. For 1D spatial models, the sets of neutrality in the space $(d,r)$ are presented in figure \ref{fig:BD1}(b). The red line corresponds to the DB process, the blue line - to the BD process, and the black line to the nonspatial Moran model. These lines become very close to the $r/d=1$ line as $N$ increases.

\section{Approximate results for fixation probability in 2D}
\label{sect:2D}

In the following, we apply some of the understanding gained from
the previous analysis on DB models and generalize it to find an approximate analytical result in higher dimensions. The 1D case was rather special due to the following properties:

\begin{itemize}
\item The number of mutant/normal neighbours at every time during the evolution of the clone remains constant on the two fronts of the mutant clone. The only two exceptions are when there is only one mutant in the system ($N-1$ normal cell) or one normal ($N-1$ mutant).
\item Due to the above property, the ratio of transition probabilities
for any clone size remains the same as for the mixed population  Moran model other than the two exception of $n=1$ and $n=N-1$. Including these two new transition probabilities into the solutions of the Kolmogorov equation lead to an exact result for the fixation probability of the 1D model.
\item The topology of the mutant clone does not change with time. If we
begin with a single mutant the domain of the mutant clone always remains a simply connected region.
\end{itemize}

\begin{figure}[!h]
\begin{center}
\epsfig{figure=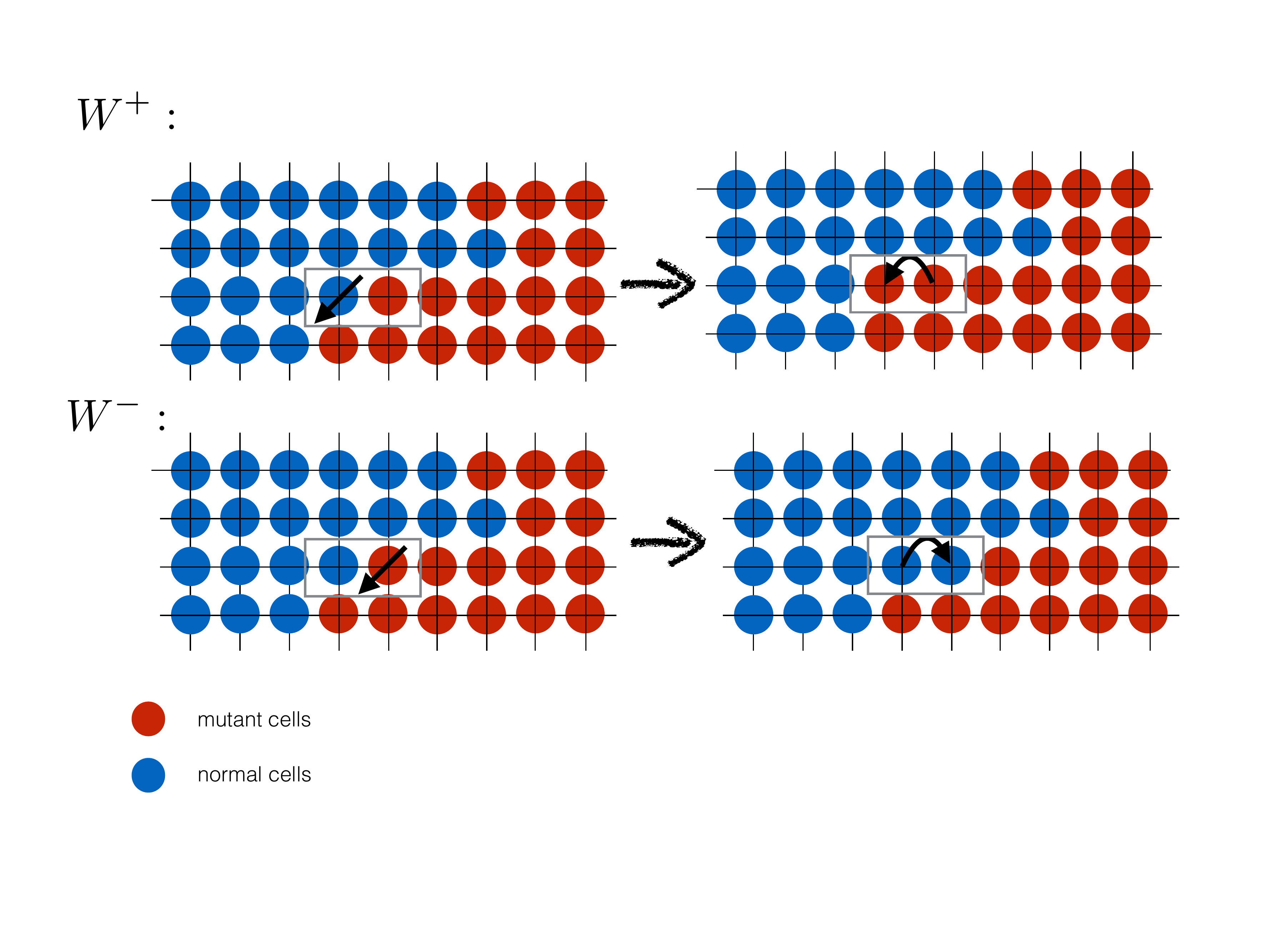, height=240pt,
width=300pt,angle=0}
\end{center}
\caption{Sequence of death and birth events on the boundary of mutant clone (red) that give rise to the increase or decrease of
mutant population. (Figure is depicted for a DB process).}
\label{voter1p}
\end{figure}

\begin{figure}[!h]
\begin{center}
\epsfig{figure=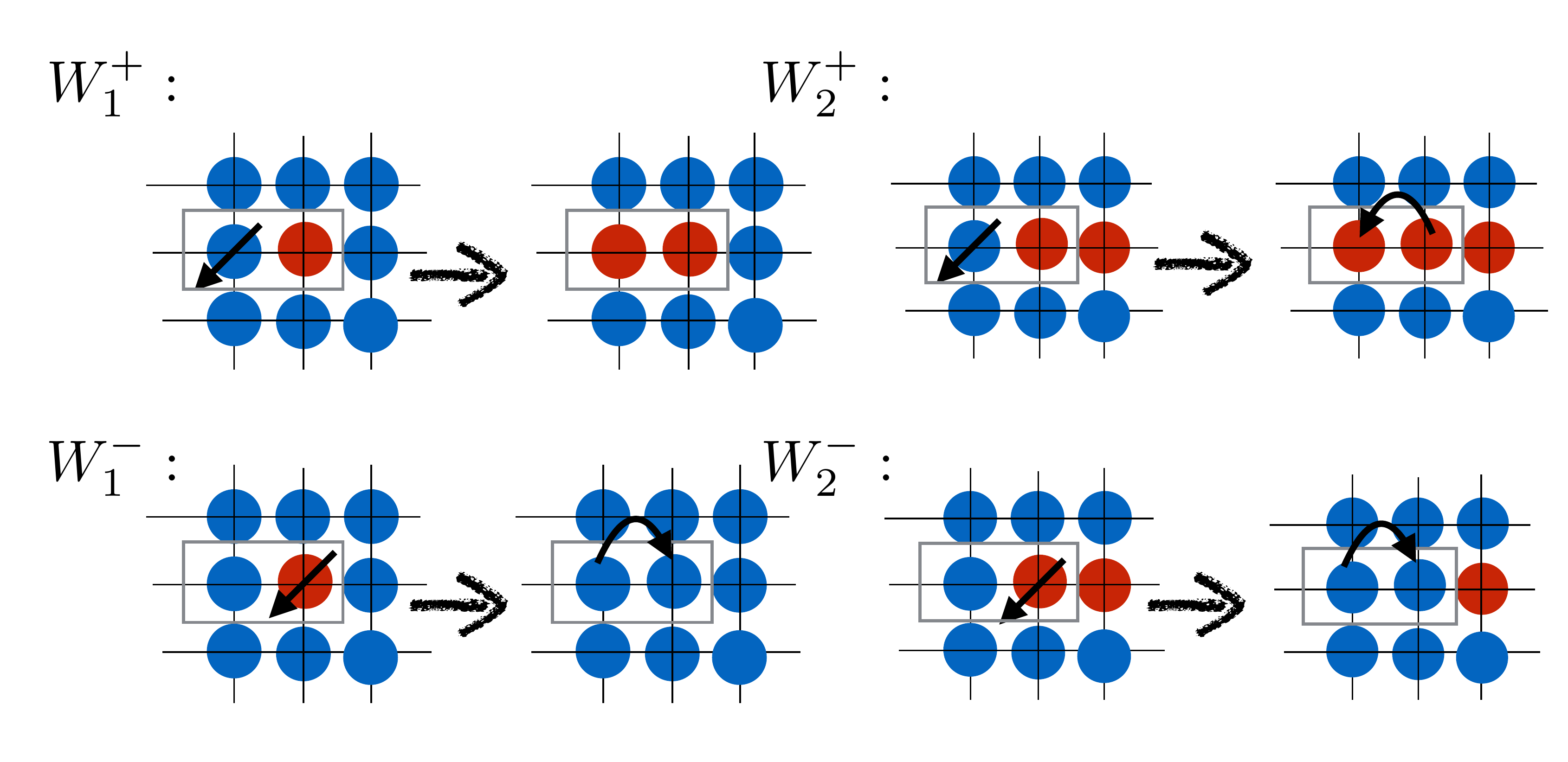, height=170pt,
width=340pt,angle=0}
\end{center}
\caption{Sequence of death and birth events in the early stages of one-mutant and two-mutant clone (red=mutant). (Figure is depicted for a DB process).}
\label{voter2}
\end{figure}

In higher dimensions, none of the above properties are correct in a strict mathematical sense. The transition probabilities in two- or higher dimensions are not only functions of the clone size but of the geometry of the clonal front. Also there are elementary events that can lead to splitting of a clone into two and thus the topology of a clone might change with time. As in the 1D case, we assume periodic boundary conditions and without loss of generality we consider a regular square lattice with degree of connectivity $k$. The sequence of events for a DB process on a square lattice is depicted in figures \ref{voter1p} and \ref{voter2}.

The ratio of transition probabilities in higher dimensions for loss or gain of a mutant cell differs for large clone sizes and small clone sizes. Here, we {\it assume} that the dominant contribution comes from $m=1$ (in 1D, $m=1$ was the only different term), and use the approximation $W^{-}_{m}/W^{+}_{m}\approx d/r$ for $m\geq 2$ (see Appendix B). We have

\bea
\frac{W^{-}_{\rm DB-2D, m}}{W^{+}_{\rm DB-2D, m}} = \displaystyle
\left\{ \begin{array}{cc}
\approx d/r &  2 < m < N-2\\
d(r + k-1)/(k r)  = \gamma^{\rm DB-2D}_{1}& m =1 \\
= d/r& m=2\\
\end{array}
\right.
\label{W2D-DB}
\eea

Thus, we obtain,

\bea
\pi^{\rm DB-2D}_{1} &\approx& \displaystyle \frac{1}{\displaystyle 1 + \sum^{N-1}_{j=2}\left(1+\prod^{j}_{k=2}\frac{W^{-}_{k}}{W^{+}_{k}}\right)\gamma^{\rm DB-2D}_{1}},\nonumber\\
&=& \displaystyle \frac{1}{\displaystyle 1 + \left( \frac{1 - \left(\frac{d}{r}\right)^{N-2}}{1 - \left(\frac{d}{r}\right)}\right)\gamma^{\rm DB-2D}_{1}},
\label{2Dfix-finN}
\eea

\begin{figure}[!h]
\begin{center}
\epsfig{figure=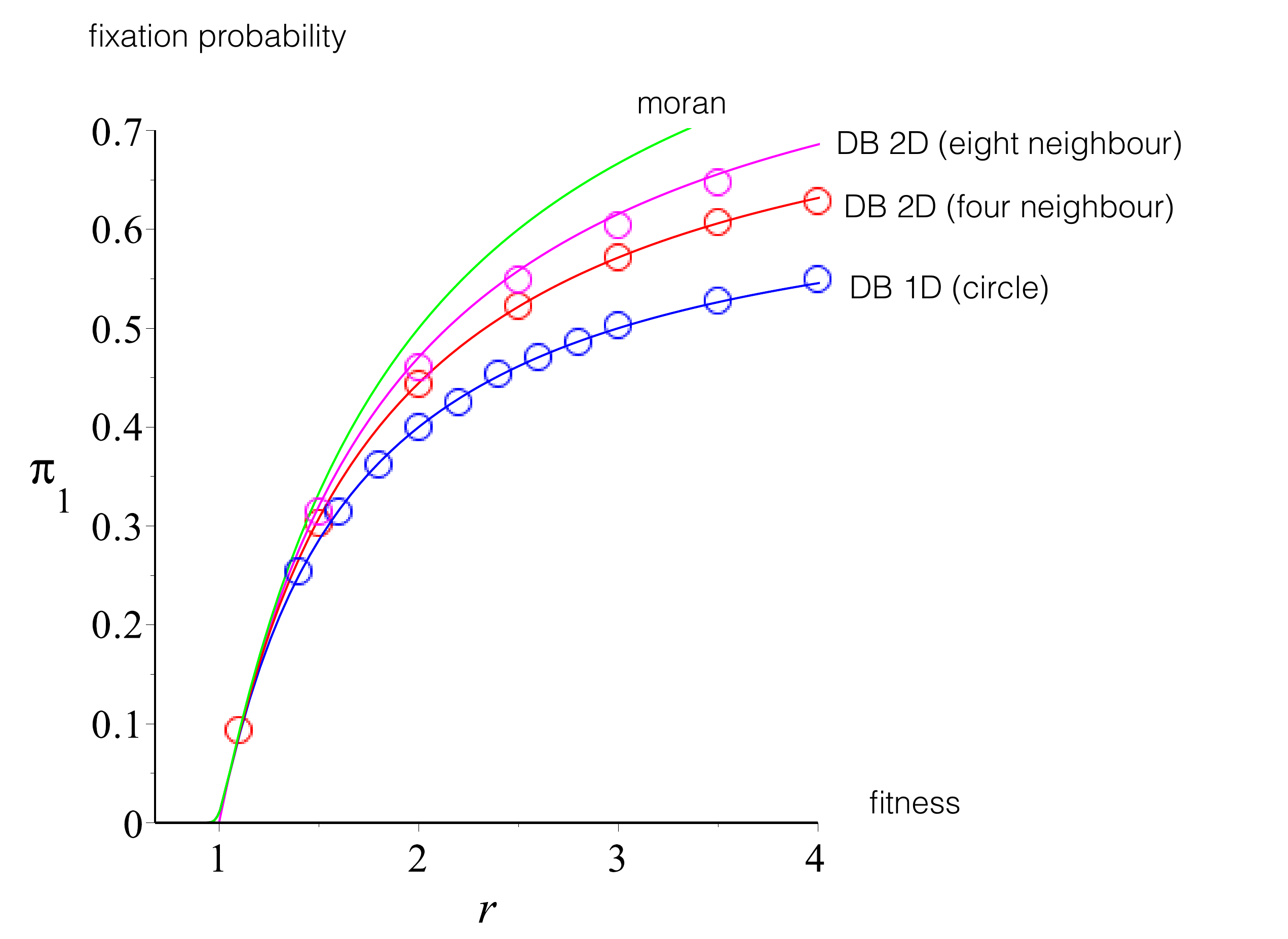, height=280pt,
width=380pt,angle=0}
\end{center}
\caption{Comparison of analytical results for the fixation probability $\pi_{1}$ as a function of the division rate $r$ for 1D DB (N=20) and 2D DB (N=20x20) cases with k=4 (4-neighbour) and k=8 (8-neighbour) regular lattices with periodic boundary conditions. The solid lines are analytical result from equation~(\ref{piA}) (mixed population Moran model), equation~(\ref{1Dfix}) (1D DB with $d=1$) and equation~(\ref{2Dfix}) (2D DB with $d=1$) while circles are the results of stochastic simulations with 40,000 iterations.}
\label{DB-simulations-graph6}
\end{figure}

\nd upon substitution from equation~(\ref{W2D-DB}) we end up with an algebraic expression for $\pi^{\rm DB-2D}_{1}$, for large $N$,

\bea
\pi^{\rm DB-2D}_{1} \approx \frac{k(r-d)}{kr+d(r-1)}.
\label{2Dfix}
\eea
For $d=1$ and $k=2$, this reduces to $2(r-1)/(3r-1)$.
For large $k$, we obtain $\pi^{\rm DB-2D}_{1} \approx 1-d/r$.
Equation~(\ref{2Dfix}) can be compared with the result of the stochastic simulations result on a square lattice ($k=4$, $d=1$) as shown in figure~\ref{DB-simulations-graph6}.

\begin{figure}[!h]
\begin{center}
\epsfig{figure=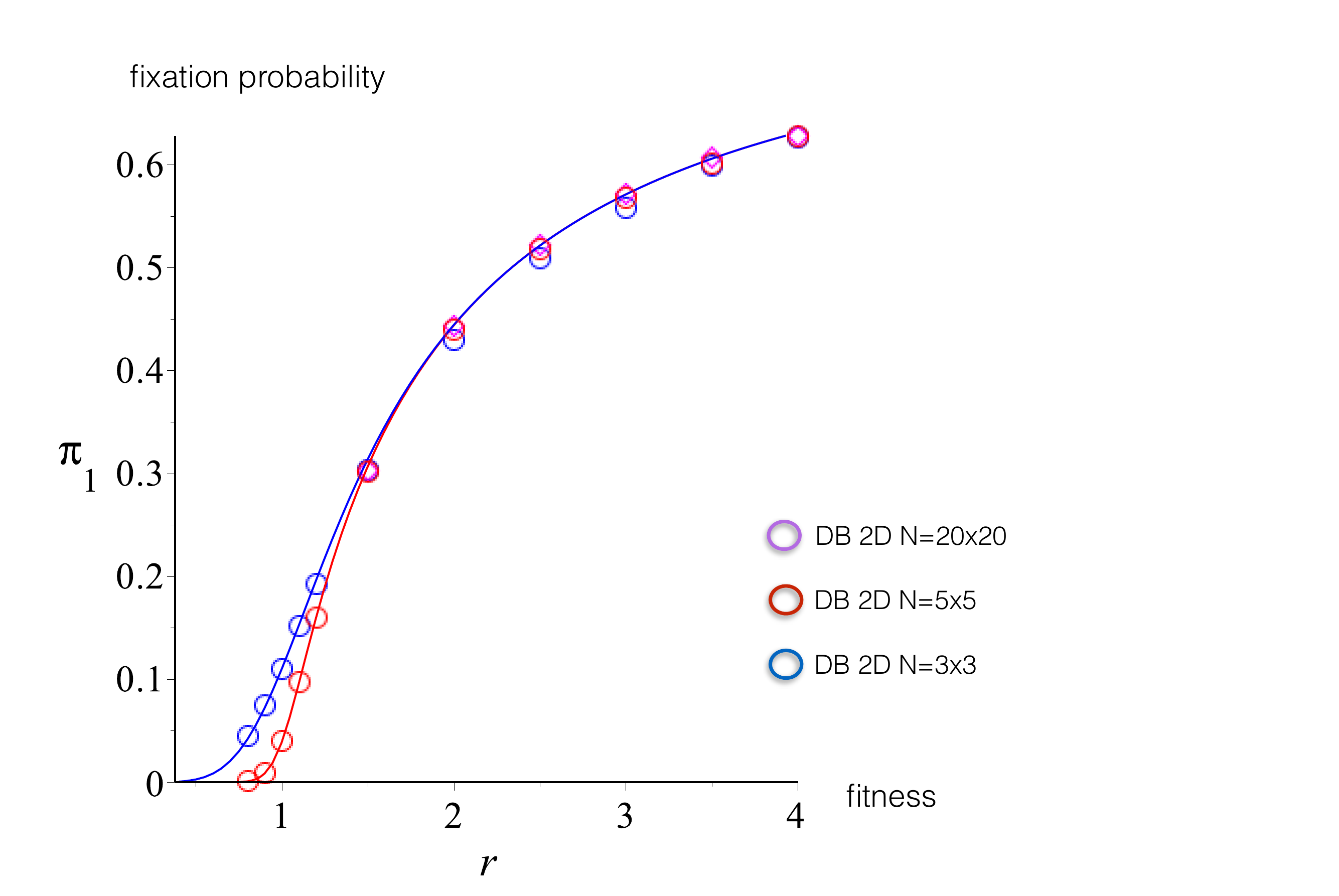, height=280pt,
width=380pt,angle=0}
\end{center}
\caption{Comparison of analytical results for the fixation probability $\pi_{1}$ as a function of the division rate $r$ for a square lattice 2D DB with different lattice sizes cases with k=4 (4-neighbour). Solid lines are analytical results from equation~(\ref{2Dfix-finN}) while circles are the results of stochastic simulations with 40,000-80,000 iterations (for smaller sizes we used higher number of iterations to reduce the statistical error).}
\label{DB-simulation-graph5}
\end{figure}

We can even test the validity of this approximation for 2D regular graphs for smaller lattice sizes. To do so we can use the finite-$N$ contribution from equation~(\ref{2Dfix-finN}). In figure~\ref{DB-simulation-graph5} we plotted the result for square lattices ($k=4$) with $N=3\times3$ and $N=5\times5$ and $N=20\times 20$. The match is good for closer to neutral limit cases while there is slight deviation from the analytical prediction as fitness increases away from the neutral limit.

\begin{figure}[!h]
\begin{center}
\epsfig{figure=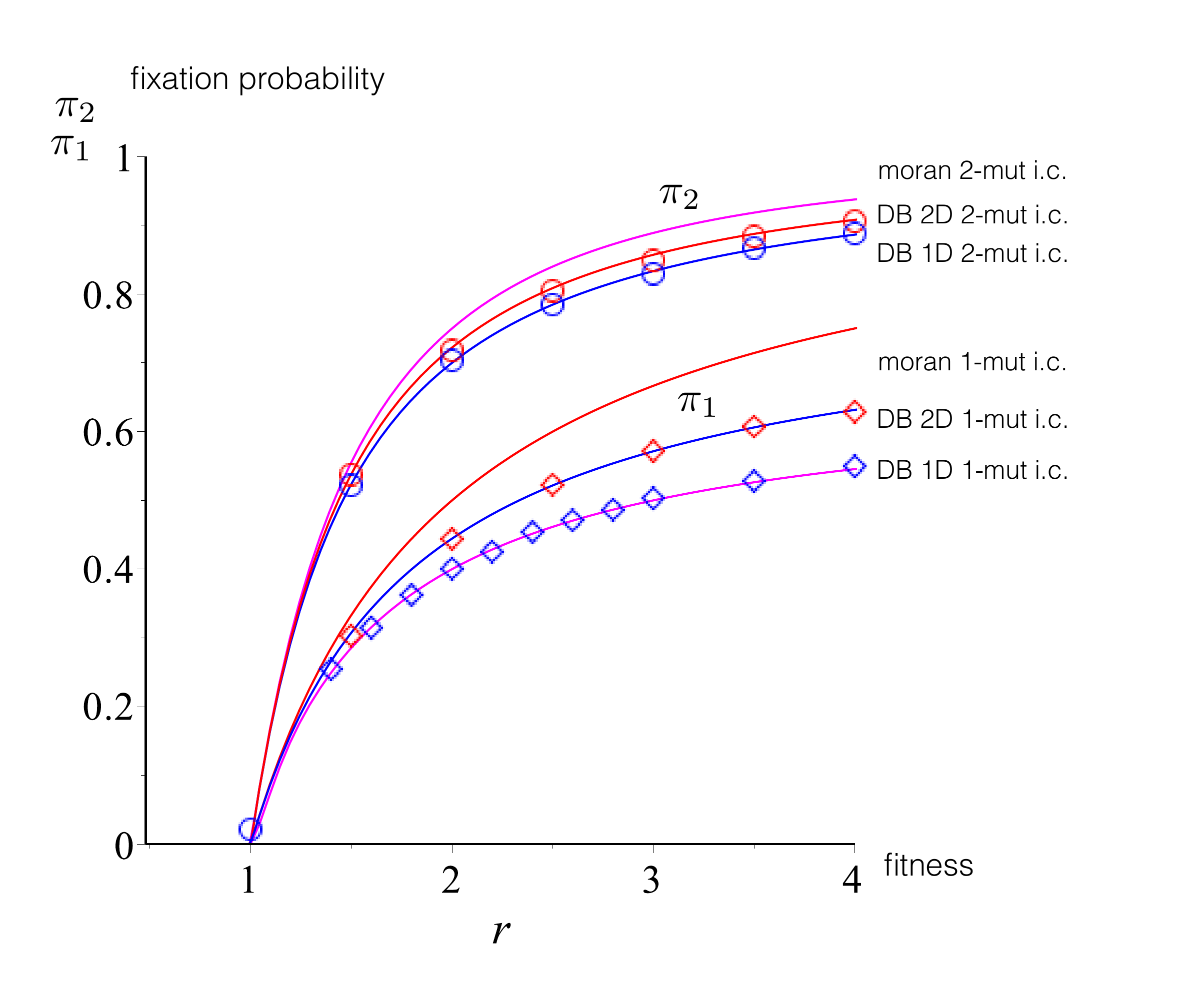, height=320pt,
width=380pt,angle=0}
\end{center}
\caption{Comparison of fixation probability for two-mutant and one-mutant initial conditions for $N=100$ (1D DB) and $N=20\times 20$ square lattice (2D DB) as a function of division rate $r$. Solid lines are corresponding theoretical results. As can be seen the two-mutant results are closer to Moran formula, equation~(\ref{moran-2mut}).}
\label{DB-simulations-2mut}
\end{figure}

We can also further test validity of equations~(\ref{2Dfix-finN}) and (\ref{2Dfix}) by comparing the results for fixation probability when one begins with {\it two} mutants, i.e. the transition probability
$\gamma_{1}$ which is claimed to be the main cause of deviation from isothermal behaviour in the DB case does not have any effect and we do expect the fixation probability to be close to a mixed population formula with the two-mutant initial condition. Notice that $\gamma_{1}$ is not the only cause of deviation from the isothermal result as other transition probabilities $\gamma_{i}$ are only approximately $l/r$. This is in fact the case for the results of one-mutant and two-mutant fixation probabilities which are depicted for a square lattice with $N=20\times 20$ and $d=1$ in figure~\ref{DB-simulations-2mut}.

Recalling the Moran fixation probability with the two-mutant initial condition $\pi_{2}$,

\be
\pi^{\rm Moran}_{2} =\frac{1 - \displaystyle\left(\frac{1}{r}\right)^{2}}{1 - \displaystyle\left(\frac{1}{r}\right)^{N}}.
\ee

The exact result can be obtained (in either 1D circle or 2D regular lattices) by using the recursive relation,

\be
\pi_{2} = \left(1 + \frac{1}{r}\right)\pi_{1},
\label{moran-2mut}
\ee

\nd which is obtained from solutions of Kolmogorov equation. This leads to the following results for $\pi_{2}$ in 1D and 2D cases in the large-$N$ limit,

\bea
\pi^{\rm DB-1D}_{2} = \frac{3r +1}{3r -1}\left(1 - \frac{1}{r}\right),\nonumber\\
\pi^{\rm DB-2D}_{2} = \frac{5r +3}{5r -1}\left(1 - \frac{1}{r}\right).
\label{2Dfix-2mut}
\eea

The above approximation fits a large-$N$ simulation results well as shown in figure~\ref{DB-simulations-2mut}. Since equations~(\ref{2Dfix-2mut}) are obtained using a mixed population form of Kolmogorov equation, the fits with the simulations further support our assumptions in deriving the formula for the 2D DB fixation probability, given by equation~(\ref{2Dfix-finN}).

In the case of a BD updating, while the special case of $d=1$ leads to the isothermal theorem for regular graphs, the fixation probability for arbitrary values of $r$ and $d$ might not follow isothermal behaviour analogous to the case discussed earlier. Similar to the 2D DB case we obtain,

\bea
\frac{W^{-}_{\rm BD-2D, m}}{W^{+}_{\rm BD-2D, m}} = \displaystyle
\left\{ \begin{array}{cc}
\approx d/r &  2 < m < N-2\\
kd/(r(d+k-1)) = \gamma^{\rm BD-2D}_{1}& m =1\\
= d/r & m=2\\
\end{array}
\right.
\label{eq17}
\eea
This gives, for large $N$,

\bea
\pi^{\rm BD-2D}_{1} \approx \frac{(d+k-1)(r-d)}{(d+k-1)(r-d)+kd}.
\label{2Dfix2}
\eea
For $d=1$, equation~(\ref{2Dfix2}) reduces to $1-1/r$ (independent of $k$). For large $k$, we have $\pi^{\rm BD-2D}_{1}\approx 1-d/r$. Setting $k=2$, this reduces to the 1D case.

\section{Discussion}
\label{sect:disc}

In this work, we discussed generalized versions of death-birth and birth-death updating models for evolutionary dynamics on graphs and spatial structures. In contrast to most previous approaches, we assume that the  mutants can differ from the wild type not only by their birth rate ($r$), but also by their death rate, ($d$), giving rise to a a two-parameter model where $r$ and $d$ are independent arbitrary parameters affecting the selection process. This is in fact more realistic, as in many scenarios either death rate or birth rate or both can vary and affect the selection dynamics in tissues in situations involving invasion mechanisms such as cancer. Focusing on regular lattices with periodic boundary conditions, we show that the models for both DB and BD updating are exactly solvable in a 1D circle.

We further studied to what extent the isothermal theorem can be generalized. This theorem states that the selection dynamics are the same on any isothermal graph as they are for a mixed population/complete graph. This theorem was proven in \cite{lieberman} to hold for the $d=1$ BD update, and here we show that it also holds for the $r=1$ DB update. We further demonstrated that the more general, two-parameter models deviate from this theorem. The BD and DB systems exhibit fixation probabilities that are different from each other and from the canonical mass-action Moran result.

While for the case of the complete graph, the difference is vanishingly small for large graphs,  it remains finite for 1D and 2D spatial structures. In general, we conjecture that the difference between the two models has the order of the inverse of the degree of connectivity of the graph. For example, the connectivity of the complete graphs (the mass-action system) is $N$, and thus the difference scales with $1/N$. The connectivity of regular spatial lattices is given by the number of neighbors of each node. In the 1D ring, the number of neighbors is $2$, and in typical 2D lattices, it is 4 or 8. Thus the difference in the fixation probability between the BD and DB models is the largest for the 1D ``nearest neighbor" ring, it is slightly smaller in the case of the von Neumann (4-cell) neighborhood in 2D, and still smaller in the case of the Moore (8-cells) neighborhood. It decays with the the number neighbors.

The reason for the deviation from the isothermal theorem is the different transition probabilities for the case when there is only one mutant in the system, compared to the rest of the transition probabilities. When the quantity of interest is the fixation probability starting from more than 1 mutant cells, then the difference between DB and BD models (and the deviation from the isothermal theorem) is smaller, compared to the case when we start with one mutant.

Depending on the parameters, the spatial DB and DB processes could be characterized by a smaller or larger mutant fixation probability compared with the conventional non-spatial Moran result. In particular, for the BD process, the fixation probability is higher than the Moran value as long as $d>1$. For the DB process, the fixation probability is higher than the Moran value as long as $r<1$.

The generalized two-parametric BD and DB models can be applied to investigate the evolutionary dynamics of tissue turnover. We would like to emphasize the importance of considering both birth and death rates of mutants (compared to those of the surrounding cells) in terms of the conceptual construct of the hallmarks of cancer \cite{hanahan}. Three of them are the most relevant for our study. They are listed below and put in our context by linking them with the cells' birth and death rates \cite{vogelstein,vogelstein2,mendelsohn}:
\begin{enumerate}
\item  {\it Self-sufficiency in growth signals}. While normal	cells cannot	proliferate	in	the	absence	of	 stimulatory	signals, cancer	cells can do this with the help of oncogenes, which	mimic	normal	growth	 signaling. This can be achieved by means of different mechanisms. For example, cells can acquire the	ability	
to	synthesize	their	own	growth	factors, e.g. the	production	of	PDGF	(platelet	derived	growth	factor)	 and	TGF-$\alpha$	(tumor	
growth	factor	$\alpha$)	by	glioblastomas	and	sarcomas. Further, cell	surface	receptors	that	transduce	 growth-stimulatory	signals	into	the	 cell,	for	 example	EGFR	and	HER2/neu,	may	be	over	 expressed	 or	structurally	altered, leading	 to	ligand-independent	signaling. Finally,  downstream	targets	of	the	 signaling	pathway	can	be	altered, e.g. the	Ras	oncogene, which
is	found	mutated	in about	25\%	of	human	tumors. In all these cases, the mutant cells are characterized by an {\bf increased birth rate} compared to the surrounding cells.
\item {\it Insensitivity to antigrowth signals.} Antigrowth	signals	can	block	proliferation	by (i)	forcing	 cells	out	of	the	
active	proliferative	cycle	into	the	quiescent	(G0
)	state,	until	appropriate	growth	signals	put	
them	back	into	the	cell	cycle;	or	(ii)	inducing	differentiation,	which	permanently	removes	
their	proliferative	potential.	Cancer	cells	evade	these	antiproliferative	signals, by e.g. loss	of	 TGF$\beta$, loss	of	Smad4, or  loss	of	 CDK	inhibitors	such as	p16,	p21,	 or p53. The corresponding cells again are characterized by an {\bf increased birth rate} compared to the surrounding cells.
\item {\it Evading apoptosis.} The	ability	of	tumor	cell	populations	to	expand	in	number	is	determined	 not	only	by	the	rate	of	cell	 proliferation	but	also	by	the	rate	of	cell	attrition.	 Programmed	cell	death (apoptosis)	represents	a	major	source	of	this	 attrition.	Resistance	
to	apoptosis	as	can	be	acquired	by	cancer	cells by e.g. loss	of	p53	(which normally	activates	 pro-apoptotic	proteins and 	represents	the	 most	common	
loss	of	a	proapoptotic	regulator), or
by activation	or	upregulation	of	anti-apoptotic	Bcl2. In these cases, the mutants are characterized by a {\bf decreased death rate} compared to the surrounding cells.
\end{enumerate}
Other applications are the cases when selection dynamics are affected by the introduction of a drug, which increases the mutant cell death while the birth rate is determined by the cell division rate and is independent of the drug concentration. This can be applied to problems in both cancer and infectious diseases, where spatial structures affect selection dynamics.

Another interesting application is the stem cell dynamics in the case of the intestinal crypt. Recently, Vermuelen et al \cite{Vermeulen2013} investigated the dynamics of niche success in the intestinal crypt base by inducing different types of mutations including, APC, p53 and Kras, and were able to measure the fraction of mutant stem cell clones at various time points for a wide number of crypts (in mice). The authors fitted the results to a simple birth-death model on a one-dimensional ring. They used Bayesian inference to infer model parameters such as proliferation rate $r$ and total niche size $N$. It has been observed that Kras oncogenic mutation infer a relatively high selection advantage in a newly introduced mutant to the stem cell niche while ${\rm APC}^{-/+}$ is weakly disadvantageous and ${\rm APC}^{-/-}$ mutation in a background of ${\rm APC}^{-/+}$ mutants is weakly advantageous. Similarly, it has been reported that p53 mutations in a normal intestinal base infer a very small selection advantage while in the inflamed gut the advantageous p53 has the higher chance of succeeding in the niche.

\begin{figure}
\begin{center}
\scalebox{0.2}{
\includegraphics*{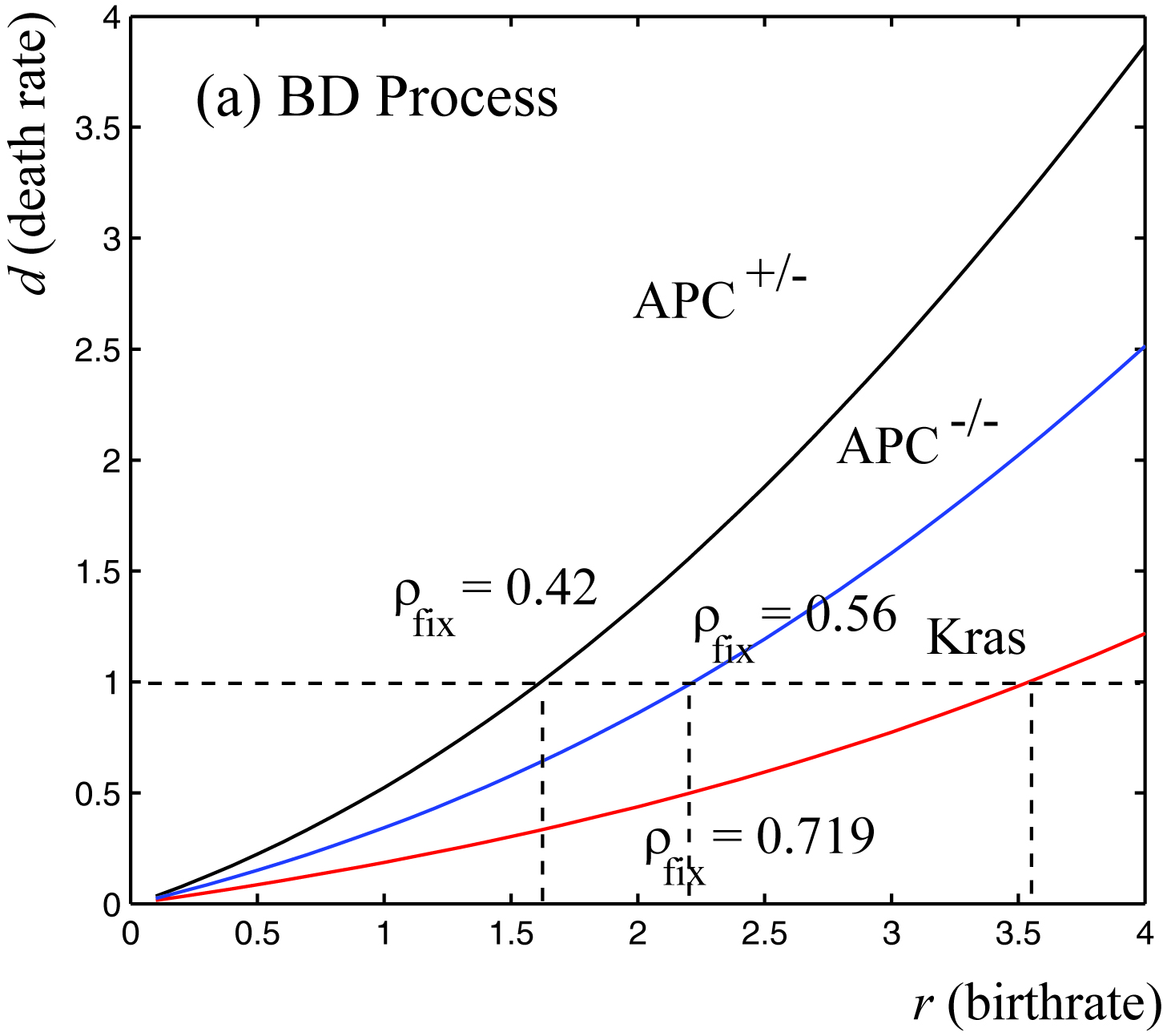}}
\scalebox{0.2}{
\includegraphics*{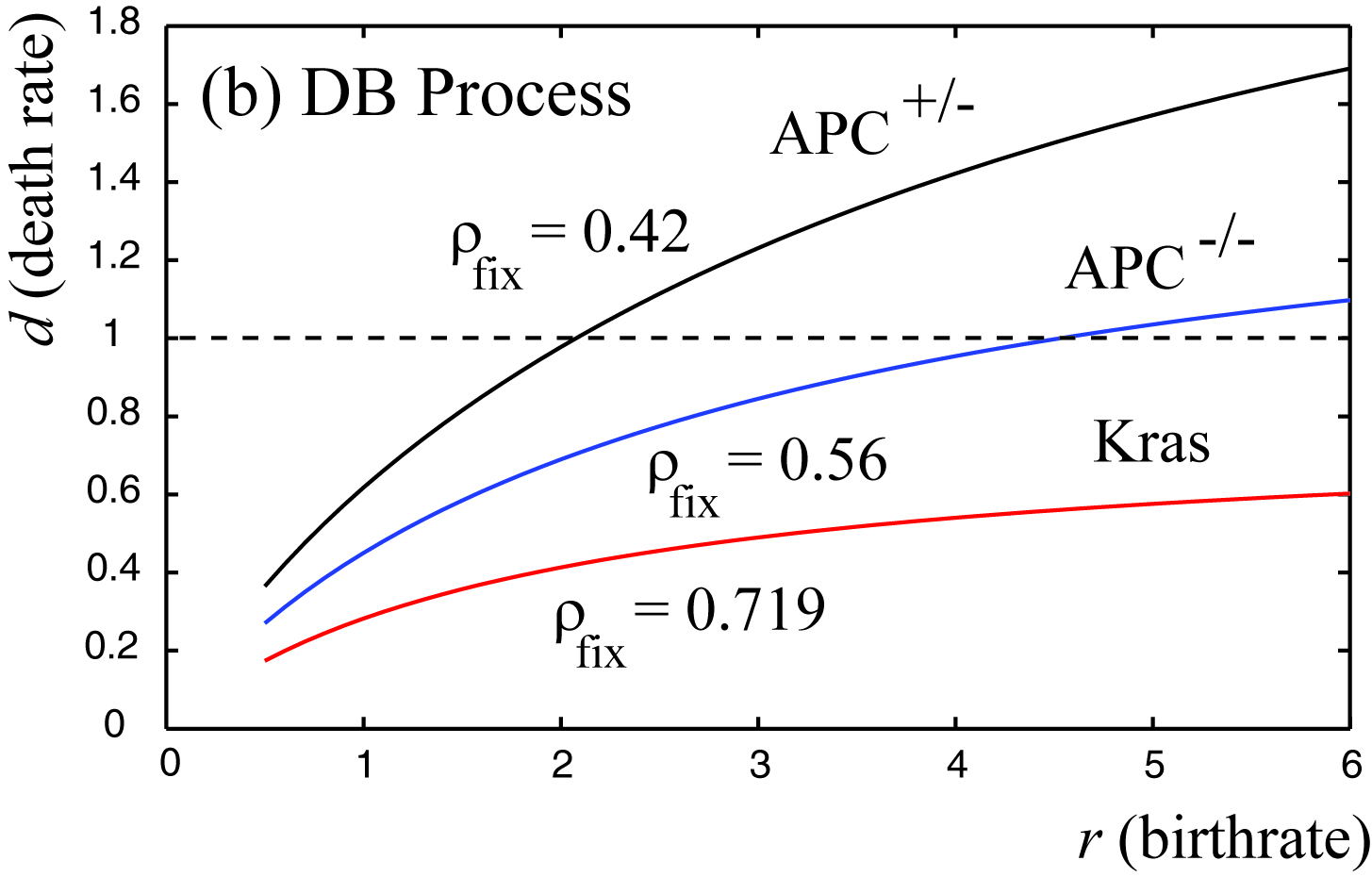}}
\end{center}
\caption{The plots of $(d,r)$ sets for different mutations. The values of the fixation probability are extracted from \cite{Vermeulen2013}.}
\label{crypt-rd}
\end{figure}

As discussed above, to compare with the experimental data, the birth and death rates of normal and mutant stem cells should be taken into account independently, particularly when both mechanisms are contributing to the selection dynamics at the same time. In the case of intestinal crypt stem cells, aside from the fact that different mutations can confer different death and birth rates (or combination of both), other mechanisms such as symmetric differentiation, and cell cycle and quiescent states can act as additional effective mechanisms for a death event in such an evolutionary model. We have used the the reported fixation probability to estimate the possible set of death and birth rates. We also applied both DB and BD processes to see if there is any significant difference between the two models. The results are depicted in figures~\ref{crypt-rd}a and~\ref{crypt-rd}b.

In figure \ref{crypt-rd}a, we have plotted the lines indicating possible sets of birth rate and death rate for three mutants, Kras, ${\rm APC}^{-/+}$ and ${\rm APC}^{-/-}$. Dashed lines indicate the values for death rates $d=1$ and various birth rates $r$ reported in \cite{Vermeulen2013}. Similar results have been depicted for the case of a DB process where the death event occurs first. In the case of relatively advantageous mutations, such as Kras, the $d=1$ case leads to an unusual high division rate. In fact higher divisor rates also point to much higher birth rates due to negative curvature of the $(r,d)$ graphs. This is basically a special case of the duality reported in the previous sections. Our finding also supports the belief that the stem cell dynamics inside a crypt niche is dominated by birth (division) events, followed by death events due to geometrical constraints in the systems. Such death events some times are referred to as a "retraction" in the biological context \cite{Vermeulen2014}.

\vspace{0.5cm}

\noindent
{\bf Acknowledgments:} We would like to thank Arne Traulsen for fruitful discussions. M. Kohandel is supported by the Natural Sciences and Engineering Research Council of Canada (NSERC, discovery grants) as well as an NSERC/CIHR Collaborative Health Research grant.

\appendix

\section{1D processes where ``self'' is included in the neighborhood network}

\label{app2}

Let us consider a one-dimensional, nearest neighbor process, where the cell itself is considered to be one of its own neighbor. Therefore, instead of two neighbors, each cell has three neighbors. Performing calculations similar to those of Section \ref{sect:1D}, we obtain the following results. Note that although qualitatively, the fixation probabilities behave similar  to those of Section \ref{sect:1D}, the numerical values differ significantly  in the two processes. Thus, whether or not we include ``self" in the neighborhood will significantly change the resulting fixation probabilities, and the change does not disappear in the large $N$ case, like it did in the mass-action scenario.

\subsection{The BD process}

We have the following expression for transition rate ratios:
\bea
\gamma_m^{BD}=\frac{W^{-}_{\rm BD, m}}{W^{+}_{\rm BD, m}} = \displaystyle
\left\{ \begin{array}{cc}
d(1+2d)/(r(d+2)) &  1 < m < N-1\\
d/r  & m = 1\\
d/r & m = N - 1
\end{array}
\right.
\label{eq11-2}
\eea
Note that if $d=1$, we have $\gamma_m^{BD}=1/r$ for all $m$. Using equation~(\ref{eq4}), the probability of fixation can be easily calculated. Because the expression is rather cumbersome, here we only present the approximation for the case when simultaneously, $r>d$, $r>d(1+2d)/(2+d)$, and $N\to\infty$:
$$\pi_1^{BD}=1-\frac{d(d+2)}{r(d+2)-d(d-1)}.$$
The isothermal theorem (for the general $N$) is satisfied for $d=1$.

\subsection{The DB process}

In this case, the transition rate ratios become:
\bea
\gamma_m^{DB}=\frac{W^{-}_{\rm DB, m}}{W^{+}_{\rm BD, m}} = \displaystyle
\left\{ \begin{array}{cc}
d(r+2)/(1+2r) &  1 < m < N-1\\
d/r  & m = 1\\
d/r & m = N - 1
\end{array}
\right.
\label{eq11-3}
\eea
If $r=1$, we have $\gamma_m^{DB}=d$ for all $m$. Using equation~(\ref{eq4}), the probability of fixation can be calculated; again, the expression is cumbersome.  Here we only present the approximation for the case when simultaneously, $r>d$, $d<r(1+2r)/(r+2)$, and $N\to\infty$:
$$\pi_1^{BD}=1-\frac{d(1+2r)}{d(r-1)+r(2r+1)}.$$
The isothermal theorem is satisfied for $r=1$.

\section{2D graph transition probability}

In this section we briefly discuss the ratio of transition probabilities for both DB and BD models when the mutant population is large. For simplicity, we assume $d_{\rm A}=d_{\rm B}=1$, $r_{\rm A}=1$, and $r_{\rm B}=r$. Using equations~(2), the ratio of transition probabilities are given by
\be
\frac{W^{-}_{\rm BD}}{W^{+}_{\rm BD}} = \displaystyle\frac{\sum_{i,j}\displaystyle\frac{ w_{ji}n_{j}(1-n_{i})}{(r-1)\sum_{k}n_{k}+N}} {\sum_{i,j}\displaystyle\frac{r w_{ji}n_{i}(1-n_{j})}{(r-1)\sum_{k}n_{k} + N}}.
\ee
For isothermal graphs the sum over indices $i$ and $j$ can be switched and the ratio becomes $1/r$ for any configuration. On the other hand, this is not quite true for the DB process. The ratio of transition probabilities, using equations~(1) are,
\be
\frac{W^{-}_{\rm DB}}{W^{+}_{\rm DB}} = \displaystyle\frac{\sum_{i,j}\displaystyle\frac{w_{ij}n_{i}(1-n_{j})}{N\left((r-1)\sum_{l}w_{il}n_{l} + k\right)}}{\sum_{i,j}\displaystyle\frac{r w_{ij}n_{j}(1-n_{i})}{N\left((r-1)\sum_{l}w_{il}n_{l} + k\right)}},
\ee
\nd where $k = \sum_{l} w_{il}$ is the number of neighbours for each site. Notice that the denominator and numerator do not simplify as the terms in the denominator now depend on $i$ as well. However, for large mutant populations, we can approximate $(r-1)\sum_{l}w_{il}n_{l} + k$ and make it independent of the index $i$. In this case, one can simplify the ratio and arrive at the result $1/r$. This is, however, incorrect for smaller populations, as discussed in section 5. The main contribution comes from the $m=1$ population. For $m =1,2$ ($m$ is the mutant population), we have
\bea
\frac{W^{-}_{1}}{W^{+}_{1}}&=& \frac{r+3}{4r}\nonumber\\
\frac{W^{-}_{2}}{W^{+}_{2}}&=& \frac{\displaystyle\frac{6}{r+3}}{\displaystyle\frac{6r}{r+3}} = \frac{1}{r}.
\label{eq14}
\eea
Similar calculations for $m=3$ gives $W^{-}_{3}/W^{+}_{3}\approx 1/r$.

\section{Generating function formalism}

\subsection{Mixed population in weak selection}

 In the following we repeat the alternative derivation of the fixation probability for a nonspatial Moran model from \cite{hooch1}. Beginning with the master equation for the Moran process in the weak-selection limit,

\bea
\frac{\partial p(n;t)}{\partial t} &=& W^{+}(n-1)p(n-1;t) + W^{-}(n+1)p(n+1;t)\nonumber\\
&-& (W^{+}(n) + W^{-}(n))p(n,t) ,
\eea

\nd where $p(n; t)$ is the probability of having $n$ mutants at time $t$ with a given initial condition. The transition rates $W^{\pm}(n)$ are given by

\bea
W^{+}(n) &=& \frac{r\cdot n(N-n)}{((rn + (N-n))(dn + (N-n))},\nonumber\\
W^{-}(n) &=& \frac{d \cdot n(N-n)}{((rn + (N-n))(dn + (N-n))},
\label{moranW}
\eea

\nd where the transition probabilities at the absorbing states are taken to be zero, $W^{\pm}(n = N) = W^{\pm}(n=0) =0$. Paper \cite{hooch1} considered the special case of the above model where $d = 1, r-1 \ll 1$. This way the denominator in transition probabilities $W^{\pm}(n)$ can be approximated by $N^{2}$. The probability generating function $F(z, t)$:

\be
F(z, t) = \sum_{m=0}z^{n}p(m,t),
\ee

\nd obeys the following equation ($d=1$):

\be
\frac{\partial F(z, t)}{\partial t} = (z - 1)(rz - 1)\frac{\partial}{\partial z} \left( N\cdot F(z,t) - z\frac{\partial F(z,t)}{\partial z}\right).
\label{pde1}
\ee

\nd Notice that equation~(\ref{pde1}) is derived using the identity,

\be
\sum_{n} n z^{n} = z\frac{\partial}{\partial z} \sum_{n}z^{n}.
\ee

Boundary and initial conditions for the PDE in equation~(\ref{pde1}), are,

\bea
F(1, t) = 1~~({\rm boundary~condition})\nonumber\\
F(z, t = 0) = z^{n_{0}}~~({\rm initial~conidian}),
\eea

\nd where $n_{0}$ is the number of mutant cells at $t = 0$. The steady state solution for $F(z,t)$ is,

\be
F(z, t \rightarrow \infty) = A + Bz^{N},
\ee

\nd where coefficients $A$ and $B$ are extinction and fixation probabilities. Applying the boundary and initial conditions and noticing that  $z^{*} = 1/r$ is a fixed point of equation~(\ref{pde1}), we can solve for the coefficient $B$ to obtain,
\be
\pi_{n_{0}} = B = \frac{\displaystyle 1 - \left(\frac{1}{r}\right)^{n_{0}}}{\displaystyle 1 - \left(\frac{1}{r}\right)^{N}}.
\ee
The trivial fixed point $z^{*} = 1$ leads to the absorption probability being unity. This corresponds to the initial condition of beginning with $N$ initial mutants. Notice that in general we do not need to find a fixed point of the PDE for the generating function but rather a point $z^{*}$ (or set of points) for which the right-hand-side of equation~(\ref{pde1}) is zero.

\subsection{Evolutionary graph models in weak selection}

The above method can be generalized to derive the fixation probability of the DB (BD) model on a graph. This method has been developed by Houchmanzadeh and Vallade in the weak selection limit of BD and DB processes in \cite{hooch} and \cite{hooch3}. We will generalize their method to account for exact solutions for arbitrary values of $r$ and $d$ in 1D.

Consider a DB model with transition probabilities,

\bea
W^{+}_{\rm DB}(n_{i}) &\equiv& P(n_{1},\ldots,n_{i},\ldots,n_{N} \rightarrow n_{1},\ldots,n_{i}+1,\dots,n_{N}) \nonumber\\
&=& \frac{1-n_{i}}{(d-1)\sum_{l}n_{l} + N}\frac{r \sum_{j}w_{ij} n_{j}}{(r-1)\sum_{k}w_{ik}n_{k} + 1}\nonumber\\
W^{-}_{\rm DB}(n_{i}) &\equiv& P(n_{1},\ldots,n_{i},\ldots,n_{N} \rightarrow n_{1},\ldots,n_{i}-1,\ldots ,n_{N})\nonumber\\
 &=& \frac{dn_{i}}{(d-1)\sum_{l}n_{l} + N}\frac{\sum_{j}w_{ij}(1 - n_{j})}{(r-1)\sum_{k}w_{ik}n_{k} + 1},
\label{eq20}
\eea

\nd master equation corresponding to the above transition probabilities is,

\bea
\frac{\partial p(\{n_{i}\},t)}{\partial t} &=& W^{+}(n_{i}-1)p(n_{i}-1,t) + W^{-}(n_{i}+1)p(n_{i}+1,t)\nonumber\\
&-& (W^{+}(n_{i}) + W^{-}(n_{i}))p(n_{i},t),
\label{eq21}
\eea
where $p(n_{i}\pm1; t) \equiv p(n_{1},\ldots,n_{i}\pm1,\ldots,n_{N})$. Introducing the probability generating function $F(\{z_{i}\}; t)$:

\be
F(z_{1},\ldots,z_{N}; t) \equiv \sum_{\{n_{k}\}=0,1}z_{0}^{n_{0}}z_{1}^{n_{1}}\cdots z_{N}^{n_{N}}p(n_{1},\ldots,n_{N};t),
\label{eq22}
\ee

\nd and using the master equation~(\ref{eq21}), an equation for the generating function can be obtained (see \cite{hooch}, and also previous section),

\bea
&&\frac{\partial F(\{z_{k}\}, t)}{\partial t} = \hat{N}^{-1}_{\rm d}\sum_{i} (z_{i} - 1)\hat{N}^{-1}_{{\rm r},i} \sum_{j}r w_{ij}\left(1-z_{i}\frac{\partial}{\partial z_{i}}\right)z_{j}\frac{\partial}{\partial z_{j}}F\nonumber\\
&&+\hat{N}^{-1}_{\rm d}\sum_{i} (z^{-1}_{i} - 1)\hat{N}^{-1}_{{\rm r},i}\sum_{j}d w_{ij}\left(1-z_{j}\frac{\partial}{\partial
z_{j}}\right)z_{i}\frac{\partial}{\partial z_{i}}F,
\label{pde}
\label{eq23}
\eea

\nd where operators $\hat{N}^{-1}_{\rm d}$ and $\hat{N}^{-1}_{\rm r}$, are inverse of $\hat{N}_{\rm d}$ and $\hat{N}_{\rm r}$,  given by,

\bea
\hat{N}_{\rm d} = (d-1)\sum_{k}z_{k}\frac{\partial}{\partial z_{k}} + N,\nonumber\\
\hat{N}_{{\rm r},i} = (r-1)\sum_{k}w_{k,i}z_{k}\frac{\partial}{\partial z_{k}} + 1.
\eea

Similar to the mixed population case, boundary and initial conditions for the spatial case are,

\bea
F(z_{1}=1,\ldots,z_{N}=1, t) = 1~~({\rm boundary~condition})\nonumber\\
F(z_{1},\ldots,z_{N}, t = 0) = z_{1}^{n_{1}^{0}}\cdots z_{N}^{n_{N}^{0}}~~({\rm initial~conidian}).
\label{eq24}
\eea

\nd Here, $n^{0}_{i}$ is the number of mutant cells at each site $i$ at $t = 0$. the system has only two absorbing state corresponding to fixation or extinction of the mutants introduced to the system indicating the following stationary form of the generating function(see \cite{hooch} for detailed discussion),

\be
F(\{z_{k}\}, t \rightarrow \infty) = A + Bz_{0}z_{1}\cdots z_{N},
\label{eq25}
\ee
where $A$ and $B$ are constants to be determined from the boundary and initial conditions. The fixed points of equation~(\ref{pde}), indicated by $\{z^{*}_{1},\cdots,z^{*}_{N}\}$ determine the fixation probability \cite{hooch},

\be
\pi_{1} = B = \frac{1 - F(z^{*}_{1},\ldots,z^{*}_{N}; t )}{1 -  z^{*}_{1}z^{*}_{2}\cdots z^{*}_{N}}.
\label{eq26}
\ee

Given the initial condition and assuming that $n^{0}_{1} = 1$, and $n^{0}_{i} = 0, i\neq 1$, the fixation probability simplifies to:

\be
\pi_{1} = B = \frac{1 - z^{*}_{1}}{1 -  z^{*}_{1}z^{*}_{2}\cdots z^{*}_{N}}.
\label{fixFP}
\ee

To find $z^{*}$'s in this limit, we put the terms in front of first derivative and second derivatives ($\partial/\partial z_{i}$) terms equal to zero, separately.  In the limit of $d=1$ and $r\ll 1$ both these operators are proportional to unity and can be dropped from the rest of the calculations in the same manner as in the previous section for mixed population. The first order derivative term gives rise to a non-linear relationship between $z_{i}$'s,

\be
\sum_{j}w_{ji}(1-z^{*}_{j}) = \frac{d}{r}\left( 1 - \frac{1}{z^{*}_{i}}\right),
\label{FPeq}
\ee

\nd where the second order terms cancel out for $z^{*}_{i}= d/r$. It is straightforward to check that equation~(\ref{FPeq}), is satisfied for $z_{i}=d/r$ in an isothermal graph. In the next part we discuss the 1D limit that also give rise to exact solutions for equation~(\ref{pde}).

\subsection{1D solutions}

In the case of a 1D circle, the operator $N_{{\rm r},i}$ takes a simple form assuming that we begin with a single mutant at site $i=1$ at $t=0$ and additionally assuming that the last normal cell to be replaced by a mutant was at some site indexed by $i=K$. This leads to following form for the $N_{{\rm r},i}$,

\bea
N_{{\rm r},i} = \displaystyle
\left\{ \begin{array}{cc}
2  & i = 1\\
2r & i = K\\
(r+1) &  i\neq 1, i\neq K\
\end{array}
\right.
\label{Nr1D}
\eea

\nd this significantly simplifies equation~(\ref{pde}) and solutions for $z^{*}_{i}$  that makes RHS of equation~(\ref{pde}) zero, can be obtained (in a similar fashion equation~(\ref{FPeq}) was derived). Assuming $\{z^{*}_{k}\} = \{ z^{*}_{1}, z^{*}_{2}=z^{*},.., z^{*}_{K-1}=z^{*},z^{*}_{K},z^{*}_{K+1}=z^{*}\cdots,z^{*}_{N}=z^{*}\}$, the equation for $z_{1}^{*}$ and $z^{*}_{K}$ can be written in terms of all other fixed points, while $z^{*}_{i} = d/r, i\neq \{1, K\}$,

\bea
\sum_{j} \frac{r}{r+1}\left(z_{j} -1\right)w_{ij} - \frac{d}{2}\left(\frac{1}{z_{1}} - 1\right) =0,\nonumber\\
\sum_{j} \frac{d}{r+1}\left(\frac{1}{z_{j}} -1\right)w_{ij} - \frac{r}{2r}\left(z_{K} - 1\right)=0,~~j\neq 1,K
\label{eq30}
\label{FPeq1D}
\eea

\nd which can be further simplified to,

\bea
\frac{2r}{d(r + 1)} \left( z^{*} - 1\right) &=& 1 - \frac{1}{z^{*}_{1}},\nonumber\\
\frac{r+1}{2d} \left( z^{*}_{K} - 1\right) &=& 1 - \frac{1}{z^{*}},~~z^{*}=d/r,
\label{eq31}
\eea

\nd leading to,

\bea
z_{1} = \frac{(r+1)d}{(r-1)d+2r}, \nonumber\\
z_{K} = \frac{2d-r+1)}{r+1} \nonumber\\
z_{i} = \frac{d}{r}~~i\neq \{1, K\}.
\eea

Substituting the above values of fixed points into equation~(\ref{fixFP}),

\bea
\pi^{\rm 1D-DB}_{1} &=& \frac{1 - z^{*}_{1}}{\displaystyle 1 -  z^{*}_{1}z^{*}_{K}(z^{*})^{N-2}},\nonumber\\
&=& \frac{2d(r-d)}{\displaystyle d(d(r-1)+2r) + \left(\frac{d}{r}\right)^{N}(r^{2}(r-1-2d))},
\label{fixFP1D}
\eea

\nd which is exactly the results for the 1D DB model obtained in section 3. An exactly similar calculation can be repeated to obtain the 1D BD fixation probability as well.

Similarly, it can be argued that for a square lattice, in the large-$N$ limit, equation~(\ref{FPeq1D}), leads to (by including appropriate $N_{r}$ for 2D case, and for $d=1$),

\be
\frac{4r}{r + 3} \left( z^{*} - 1\right) = 1 - \frac{1}{z^{*}_{0}},
\label{eq33}
\ee

\nd leading to the result for the fixation probability given by equation~(\ref{2Dfix}), with $k=4$,

\be
\pi^{\rm 2D}_{1} = 1 - z^{*,{\rm 2D}}_{1} = \frac{4(r-1)}{5r -1}.
\label{eq34s}
\ee

\bibliographystyle{ieeetr}

\bibliography{references}

\end{document}